\documentclass[prd,preprintnumbers,showpacs,12pt]{revtex4}
\usepackage{epsfig}
\usepackage{graphicx}
\usepackage {axodraw}

\usepackage{bm}
\usepackage{amsmath}
\usepackage{amssymb}
\usepackage{amscd}
\usepackage{latexsym}

\newcommand{\be}{\begin{equation}}

\newcommand{\ee}{\end{equation}}
\newcommand{\bea}{\begin{eqnarray}}
\newcommand{\eea}{\end{eqnarray}}

\newcommand{\nn}{\nonumber}
\newcommand{\de}{\partial}
\def\Black{}
 \def\AliasBlue{}
 \def\Blue{}
 \def\Brown{}

\begin{document}

\newcommand{\bra}[1]{\langle #1|}
\newcommand{\ket}[1]{|#1\rangle}
\newcommand{\braket}[2]{\langle #1|#2\rangle}
\newcommand{\tr}{\textrm{Tr}}
\newcommand{\lag}{\mathcal{L}}
\newcommand{\mbf}[1]{\mathbf{#1}}
\newcommand{\desl}{\slashed{\partial}}
\newcommand{\Desl}{\slashed{D}}

\renewcommand{\bottomfraction}{0.7}
\newcommand{\epsi}{\varepsilon}

\newcommand{\nl}{\nonumber \\}
\newcommand{\tc}[1]{\textcolor{#1}}
\newcommand{\sla}{\not \!}
\newcommand{\spinor}[1]{\left< #1 \right>}
\newcommand{\cspinor}[1]{\left< #1 \right>^*}
\newcommand{\Log}[1]{\log \left( #1\right) }
\newcommand{\Logq}[1]{\log^2 \left( #1\right) }
\newcommand{\mr}[1]{\mathrm{#1}}
\newcommand{\cw}{c_\mathrm{w}}
\newcommand{\sw}{s_\mathrm{w}}
\newcommand{\ct}{c_\theta}
\newcommand{\st}{s_\theta}
\newcommand{\gt}{{\tilde g}}
\newcommand{\gtp}{{{\tilde g}^\prime}}
\renewcommand{\i}{\mathrm{i}}
\renewcommand{\Re}{\mathrm{Re}}
\newcommand{\yText}[3]{\rText(#1,#2)[][l]{#3}}
\newcommand{\xText}[3]{\put(#1,#2){#3}}


\def\to{\rightarrow}
\def\ptl{\partial}
\def\beq{\begin{equation}}
\def\eeq{\end{equation}}
\def\bea{\begin{eqnarray}}
\def\eea{\end{eqnarray}}
\def\nn{\nonumber}
\def\half{{1\over 2}}
\def\rhalf{{1\over \sqrt 2}}
\def\calo{{\cal O}}
\def\call{{\cal L}}
\def\calm{{\cal M}}
\def\del{\delta}
\def\eps{\epsilon}
\def\lam{\lambda}

\def\anti{\overline}
\def\delfac{\sqrt{{2(\del-1)\over 3(\del+2)}}}
\def\heff{h'}
\def\square{\boxxit{0.4pt}{\fillboxx{7pt}{7pt}}\hspace*{1pt}}
    \def\boxxit#1#2{\vbox{\hrule height #1 \hbox {\vrule width #1
             \vbox{#2}\vrule width #1 }\hrule height #1 } }
    \def\fillboxx#1#2{\hbox to #1{\vbox to #2{\vfil}\hfil}   }

\def\braket#1#2{\langle #1| #2\rangle}
\def\gev{~{\rm GeV}}
\def\gam{\gamma}
\def\sn{s_{\vec n}}
\def\sm{s_{\vec m}}
\def\mm{m_{\vec m}}
\def\mn{m_{\vec n}}
\def\mh{m_h}
\def\sumn{\sum_{\vec n>0}}
\def\summ{\sum_{\vec m>0}}
\def\vl{\vec l}
\def\vk{\vec k}
\def\ml{m_{\vl}}
\def\mk{m_{\vk}}
\def\gp{g'}
\def\gt{\tilde g}
\def\hw{{\hat W}}
\def\hz{{\hat Z}}
\def\ha{{\hat A}}

\def\yy{{\cal Y}_\mu}
\def\yyt{{\tilde{\cal Y}}_\mu}
\def\lq{\left [}
\def\rq{\right ]}
\def\dmu{\partial_\mu}
\def\dnu{\partial_\nu}
\def\dmus{\partial^\mu}
\def\dnus{\partial^\nu}
\def\gp{g'}
\def\gpt{{\tilde g'}}
\def\ggs{\frac{g}{\gs}}
\def\eps{{\epsilon}}
\def\tr{{\rm {tr}}}
\def\V{{\bf{V}}}
\def\W{{\bf{W}}}
\def\Wt{\tilde{ {W}}}
\def\Y{{\bf{Y}}}
\def\Yt{\tilde{ {Y}}}
\def\L{{\cal L}}
\def\s{s_\theta}
\def\st{s_{\tilde\theta}}
\def\c{c_\theta}
\def\ct{c_{\tilde\theta}}
\def\gt{\tilde g}
\def\et{\tilde e}
\def\At{\tilde A}
\def\Zt{\tilde Z}
\def\Wpt{{\tilde W}^+}
\def\Wmt{{\tilde W}^-}

\newcommand{\Apt}{{\tilde A}_1^+}
\newcommand{\Bpt}{{\tilde A}_2^+}
\newcommand{\Amt}{{\tilde A}_1^-}
\newcommand{\Bmt}{{\tilde A}_2^-}
\newcommand{\Wtp}{{\tilde W}^+}
\newcommand{\Atp}{{\tilde A}_1^+}
\newcommand{\Btp}{{\tilde A}_2^+}
\newcommand{\Atm}{{\tilde A}_1^-}
\newcommand{\Btm}{{\tilde A}_2^-}
\def\mathswitchr#1{\relax\ifmmode{\mathrm{#1}}\else$\mathrm{#1}$\fi}
\newcommand{\Pe}{\mathswitchr e}
\newcommand{\Pp}{\mathswitchr {p}}
\newcommand{\PZ}{\mathswitchr Z}
\newcommand{\PW}{\mathswitchr W}
\newcommand{\PD}{\mathswitchr D}
\newcommand{\PU}{\mathswitchr U}
\newcommand{\PQ}{\mathswitchr Q}
\newcommand{\Pd}{\mathswitchr d}
\newcommand{\Pu}{\mathswitchr u}
\newcommand{\Ps}{\mathswitchr s}
\newcommand{\Pc}{\mathswitchr c}
\newcommand{\Pt}{\mathswitchr t}
\newcommand{\rd}{{\mathrm{d}}}
\newcommand{\GW}{\Gamma_{\PW}}
\newcommand{\GZ}{\Gamma_{\PZ}}
\newcommand{\GeV}{\unskip\,\mathrm{GeV}}
\newcommand{\MeV}{\unskip\,\mathrm{MeV}}
\newcommand{\TeV}{\unskip\,\mathrm{TeV}}
\newcommand{\fba}{\unskip\,\mathrm{fb}}
\newcommand{\pba}{\unskip\,\mathrm{pb}}
\newcommand{\nba}{\unskip\,\mathrm{nb}}
\newcommand{\PT}{P_{\mathrm{T}}}
\newcommand{\PTmiss}{P_{\mathrm{T}}^{\mathrm{miss}}}
\newcommand{\CM}{\mathrm{CM}}
\newcommand{\inv}{\mathrm{inv}}
\newcommand{\sig}{\mathrm{sig}}
\newcommand{\tot}{\mathrm{tot}}
\newcommand{\backg}{\mathrm{backg}}
\newcommand{\evt}{\mathrm{evt}}
\def\mathswitch#1{\relax\ifmmode#1\else$#1$\fi}
\newcommand{\M}{\mathswitch {M}}
\newcommand{\R}{\mathswitch {R}}
\newcommand{\TEV}{\mathswitch {TEV}}
\newcommand{\LHC}{\mathswitch {LHC}}
\newcommand{\MW}{\mathswitch {M_\PW}}
\newcommand{\MZ}{\mathswitch {M_\PZ}}
\newcommand{\Mt}{\mathswitch {M_\Pt}}
\newcommand{\gs}{{g''}^2}
\def\lmu{{\bf L}_\mu}
\def\rmu{{\bf R}_\mu}
\def\si{\sigma}
\def\beqar{\begin{eqnarray}}
\def\eeqar{\end{eqnarray}}
\def\refeq#1{\mbox{(\ref{#1})}}
\def\reffi#1{\mbox{Fig.~\ref{#1}}}
\def\reffis#1{\mbox{Figs.~\ref{#1}}}
\def\refta#1{\mbox{Table~\ref{#1}}}
\def\reftas#1{\mbox{Tables~\ref{#1}}}
\def\refse#1{\mbox{Sect.~\ref{#1}}}
\def\refses#1{\mbox{Sects.~\ref{#1}}}
\def\refapps#1{\mbox{Apps.~\ref{#1}}}
\def\refapp#1{\mbox{App.~\ref{#1}}}
\def\citere#1{\mbox{Ref.~\cite{#1}}}
\def\citeres#1{\mbox{Refs.~\cite{#1}}}

\def\Black{}
 \def\AliasBlue{}
 \def\Blue{}
 \def\Brown{}

\begin{flushright}
SHEP-12-20
\end{flushright}

\title{Charged di-boson production at the LHC \\
in a 4--site model with a composite Higgs boson}

 \author{Elena Accomando, Luca Fedeli and Stefano Moretti}%
 \email{e.accomando@soton.ac.uk; l.fedeli@soton.ac.uk; s.moretti@soton.ac.uk}
 \affiliation{NExT Institute and School of Physics and Astronomy, University of
 Southampton, Highfield,
 Southampton SO17 1BJ, UK}%
\author{Stefania De Curtis}%
 \email{decurtis@fi.infn.it}
 \affiliation{Istituto Nazionale di Fisica Nucleare, Sezione di Firenze,  Via G. Sansone 1, 50019 Sesto Fiorentino, Italy}%
 \author{Daniele Dominici}%
 \email{dominici@fi.infn.it}
 \affiliation{Universit\`a degli Studi di Firenze, Dip. di
 Fisica e Astronomia, 
Via G. Sansone 1, 50019 Sesto Fiorentino, Italy}%

\begin{abstract}
\noindent
We investigate the scope of the LHC in probing the parameter space of a 4-site model supplemented
by one composite Higgs state, assuming all past, current and future energy and luminosity stages of the CERN 
machine. We concentrate on the yield of charged di-boson production giving two opposite-charge different-flavour 
leptons and missing  (transverse) energy, i.e., events induced via the subprocess $q\bar q\to  e^+\nu_e \mu^-\bar\nu_\mu$ +  ${\rm{c.c.}}$, which enables the production in the intermediate
step of all additional neutral and charged gauge bosons belonging to the spectrum 
of this model, some of which in resonant topologies.
We find this channel accessible over the background at all  LHC configurations 
after a dedicated cut-based analysis. 
We finally compare the yield of the 
di-boson mode to that of Drell-Yan processes and
establish that they have complementary strengths, 
one covering regions of parameter space precluded to the others and vice versa.
\end{abstract}

\vspace*{-1.0truecm}
\maketitle

\section{Introduction}
A strong Electro-Weak (EW) sector is expected to produce a variety of bound states including particles of spin zero and
spin one, like the $\sigma$,  the $\rho$ and the $a_1$ emerging from quark states within Quantum Chromo-Dynamics (QCD). 
Just like in QCD, the phenomenology below the scale of the  strong EW
interactions producing similar resonances can be studied in terms of an effective Lagrangian
containing these additional degrees of freedom,  based on the observed symmetries of the EW sector. 
Effective terms adding to the chiral Lagrangian just a simple scalar state or a scalar and a vector state have been recently suggested 
\cite{Contino:2010mh,Espinosa:2012ir,Bellazzini:2012tv,Gillioz:2012se}. These formulations are useful because they allow for a general
 parameterisation of the (strongly) broken symmetry of the EW sector.
These new resonances also appear in five-dimensional extensions of the Standard Model (SM) as Kaluza-Klein (KK) excitations of the SM gauge bosons \cite{Agashe:2003zs,Csaki:2003dt,Csaki:2003zu,Barbieri:2003pr,Cacciapaglia:2004zv,
Cacciapaglia:2004jz}. When deconstructed  \cite{ArkaniHamed:2001ca,Arkani-Hamed:2001nc,Hill:2000mu,Cheng:2001vd,
Abe:2002rj,Falkowski:2002cm,Randall:2002qr,Son:2003et,deBlas:2006fz} these theories emerge as gauge theories with extended $SU(2)$ symmetries. Simple four-dimensional models, 
 like the 3-site \cite{Chivukula:2006cg},   the 4-site \cite{Accomando:2008jh} and the effective composite Higgs model  \cite{Contino:2006nn} can be used to characterise the main features of the emerging phenomenology.

In its original formulation, the 4-site model describes in an effective way the interactions of extra spin-one resonances as gauge fields of a $SU(2)\otimes SU(2)$ extra gauge group. They can be thought of as the first KK excitations emerging from a five-dimensional formulation, and, due to the Anti-de Sitter/Conformal Field Theory
(AdS/CFT) correspondence, they are composite states of a strong dynamics also responsible for the breaking of the EW symmetry. 
As stated before, a strong EW sector is expected to produce also new scalar and fermion particles as bound states. In this note we consider the inclusion of a new scalar field, singlet under the gauge group, in order to reproduce in our effective description, the scalar particle recently detected by the ATLAS and CMS experiments at the Large Hadron Collider (LHC) \cite{incandela,gianotti}. The couplings of our composite Higgs particle to the SM and extra gauge bosons are free parameters for which we will derive bounds due to the EW precision tests and the present LHC measurements, as well as 
theoretical constraints enforced by perturbative unitarity requirement.

It is the purpose of this paper to investigate, in the context of the 4-site model with one composite Higgs 
state, the phenomenology of charged di-boson production at the LHC, yielding opposite-charge different-flavour lepton pairs and missing transverse energy, i.e., the process
\begin{equation}\label{eq:process}
pp(q\bar q)\to W^+W^-\to e^+\nu_e \mu^-\bar\nu_\mu~+~{\rm{c.c.}}\to e^\pm\mu^\mp E_{T}^{\rm miss}
\end{equation}
wherein the symbol $W^\pm$ refers to any possible charged spin 1 massive gauge bosons present in the model, which also allows for the production of intermediate
neutral spin 1 massless (i.e., the photon) and massive gauge bosons. In fact, having 
inserted a light composite Higgs state in the 4-site model, 
one also ought to
investigate the yield of the process
\begin{equation}\label{eq:Higgs}
pp(gg)\to h\to W^+W^-\to e^+\nu_e \mu^-\bar\nu_\mu~+~{\rm{c.c.}}\to e^\pm\mu^\mp 
E_{T}^{\rm miss}
\end{equation}
where, however, having fixed $m_h=125$ GeV 
(to account for the recent LHC results), implies that the charged gauge bosons produced in intermediate stages can only be the SM ones. 

In performing our analysis, we will take into account experimental constraints from
EW Precision Test (EWPT) data produced at LEP, SLC and Tevatron as well as experimental limits from direct 
searches of Higgs (as mentioned) and new gauge bosons performed
at Tevatron and LHC via Drell-Yan (DY) channels. In the attempt to extract a signal of the model, we will focus our attention to all energy and luminosity stages covered already or 
still foreseen for
the CERN machine. Ultimately, we will want to contrast the discovery potential of the LHC of charged di-boson production events with that of DY events, 
building on
previous studies of some of us.

The plan of the paper is as follows. In the next section we recall the details of the construction of  the 4-site model and its relation with the general  effective description of vector and axial-vector resonances. In this framework, the inclusion of a singlet composite scalar state  is straightforward. We then describe the parameter space of the model and derive both theoretical and experimental bounds constraining it. 
Sect. III will instead be devoted to describe the production and decay dynamics 
of processes 
(\ref{eq:process})--(\ref{eq:Higgs}), eventually extracting from these exclusion and evidence/discovery limits over the surviving parameter space. A final section will be devoted to
summarise our work and conclude on the comparison of the relative yields of DY and di-boson processes.

\section{The 4-site model with a singlet composite scalar state}

The 4-site model is a moose model based on the $SU(2)_L \otimes  SU(2)_1\otimes SU(2)_2 \otimes U(1)_Y$ gauge symmetry and contains three non-linear $\sigma$-model fields interacting with the gauge fields, which trigger spontaneous 
EW Symmetry Breaking
(EWSB). Its construction is presented in \cite {Accomando:2008jh} while some of its phenomenological consequences are analysed in \cite{Accomando:2010ir,Accomando:2011vt,Accomando:2011xi,Accomando:2011eu}.

In order to extend the 4-site model to include a new singlet scalar field, let us start by briefly reviewing
 its relation with the general $SU(2)_L\otimes SU(2)_R$ invariant Lagrangian describing  vector and
axial-vector resonances.
 Vector and axial-vector
resonances,
interacting with the SM gauge vector bosons, can be introduced
as in \cite{Casalbuoni:1988xm}, by assuming, in addition to the standard
global symmetry $SU(2)_L\otimes SU(2)_R$, a local symmetry
$SU(2)$ for each new vector
resonance. This symmetry group $G\otimes H$, with
\be
G=[SU(2)_L\otimes SU(2)_R]_{\rm global},
~~~H=[SU(2)_L\otimes SU(2)_R]_{\rm local},
\ee
spontaneously broken down to the custodial $SU(2)$.
Further, we can add to this sector a singlet under the symmetry describing
a possible composite Higgs state.

 The methods to
construct such a Lagrangian are the standard ones used to build
up non-linear realisations (see Refs. \cite{Coleman:1969sm,Callan:1969sn,Bando:1987br}). The necessary Goldstone bosons are
 described by three independent $SU(2)$
elements: $L$, $R$ and $M$, whose transformation properties
with respect to $G\otimes H$  are the following
\bea
&&L'(x)= g_L L(x) h_L(x),~~~R'(x)= g_R R(x) h_R(x),\nn
\eea
\be
M'(x)= h_R(x)^\dagger M(x) h_L(x),
\ee
where $g_{L,R}\in G$ and $h_{L,R}\in H$. 
These properties are very reminiscent of the linear moose field transformations \cite{Casalbuoni:2004id}.
Beside the invariance
under $G\otimes H$, we will also require an invariance under the
following
discrete left-right symmetry, denoted by $P$,
$
L\leftrightarrow R,\,M\leftrightarrow M^\dagger,
$
which ensures that the low-energy theory is parity conserving.

The vector and axial-vector resonances are introduced as linear
combinations of the gauge particles associated to the local group
$H$. The most general $G\otimes H\otimes P$ invariant Lagrangian
is given by \cite{Casalbuoni:1988xm}
\be
{\cal L}_R={\cal L}_G+{\cal L}_{kin},
\label{1.4}
\ee
where
\be
{\cal L}_G=-\frac{v^2}{4}f(\lmu,\rmu),
\label{1.5}
\ee
with
\be
f(\lmu,\rmu)= a I_1 + b I_2 + c I_3 + d I_4
\label{effe},
\ee
\bea
I_1=tr[(V_0-V_1-V_2)^2],&&I_2=tr[(V_0+V_2)^2],\nn\\
I_3=tr[(V_0-V_2)^2],&&I_4=tr[V_1^2],
\eea
and
\bea
V_0^\mu&=&L^\dagger D^\mu L,\nn\\
V_1^\mu&=&M^\dagger D^\mu M,\nn\\
V_2^\mu&=&M^\dagger(R^\dagger D^\mu R)M.
\eea
The parameters $a$, $b$, $c$, $d$ are not all independent
and are fixed so that, when decoupling the new resonances,  one recovers the
SM:
\be
a+\frac {cd}{c+d}=1.
\label{smlimit}
\ee
The covariant derivatives are defined by
\bea
D_\mu L&=&\partial_\mu L -L \lmu,\nn\\
D_\mu R&=&\partial_\mu R -R \rmu,\nn\\
D_\mu M&=&\partial_\mu M -M \lmu+\rmu M,
\eea
where $\lmu =i g"/\sqrt{2} \tau^a/2 L_\mu^a$ and $\rmu =i g"/\sqrt{2} \tau^a/2 R_\mu^a$ are the gauge fields associated to
the local symmetry group $H$.
The quantities $V_i^\mu~~(i=0,1,2)$ are, by construction,
invariant under the global symmetry $G$ and covariant under
the gauge group $H$,
\be
(V_i^\mu)'=h_L^\dagger V_i^\mu h_L,
\ee
while their transformation properties under the parity
operation, $P$, are:
\be (V_0\pm V_2)\to \pm M(V_0\pm V_2)M^\dagger,~~~~V_1\to - MV_1
M^\dagger.
\ee
Out of the $V_i^\mu$'s one can build six independent quadratic
invariants, which reduce to the four $I_i$'s listed above, when parity
is enforced.
The kinetic part for the vector fields (${\cal L}_{kin}$  in eq. (\ref{1.4})) is written in the standard form.

The 4-site model corresponds to the particular choice:
\be
a=0,\,\,\,b=c=\frac {2 f_1^2}{v^2},\,\,\,d=\frac {4 f_2^2}{v^2},\,\,\,g''=\frac {g_1}{\sqrt{2}}
\label{choice}
\ee 
and to the following identification of the chiral fields:
\be
\Sigma_1=L,\,\,\Sigma_2=M^\dagger,\,\,\Sigma_3=R.
\ee
Therefore the Lagrangian for the sector of spin 1 particles of the 4-site model, is given by:
\bea
{\cal L}_G&=&-\frac{v^2}{4}\left [\frac {2 f_1^2}{v^2}(I_2+I_3)+  \frac {4 f_2^2}{v^2}I_4\right ]\nn\\
&=&
f_1^2[D_\mu \Sigma_1)^\dagger D^\mu \Sigma_1+(D_\mu \Sigma_3)^\dagger D^\mu\Sigma_3]
+f_2^2(D_\mu \Sigma_2)^\dagger D^\mu \Sigma_2 \nn\\
&=& \sum_{i=1}^3 f_i^2(D_\mu \Sigma_i)^\dagger D^\mu \Sigma_i
\label{foursitelag}
\eea
with
\bea
&D_\mu\Sigma_1=\de_\mu\Sigma_1-i\gt \Wt_\mu\Sigma_1+i\Sigma_1 g_1
\tilde{A}_\mu^1,&\nn\\
&D_\mu\Sigma_2=\de_\mu\Sigma_2-ig_{1}\tilde{A}_\mu^{1}\Sigma_2+i\Sigma_2
g_1 \tilde{A}_\mu^2,&\,\,\,\,\,\,\,\nn\\
&D_\mu\Sigma_{3}=\de_\mu\Sigma_{3}-ig_{1}\tilde{A}_\mu^{2}\Sigma_{3}+i
\gpt\Sigma_{3}\Yt_\mu,& \label{covderivative} \eea
 where
$\tilde{A}_\mu^i=\tilde{A}_\mu^{ia}\tau^a/2$ and $g_1$ are the gauge
fields and couplings,  $\Wt_\mu=\Wt_\mu^{a}\tau^a/2$,
$\Yt_\mu=\yyt\tau^3/2$ and $\gt$, $\gpt$ are the gauge fields and
couplings  associated to $SU(2)_L$ and $U(1)_Y$, respectively.
We  have also taken  into account that the $P$ symmetry implies $f_1=f_3$.
The condition (\ref{smlimit}) is equivalent to:
\be
\frac 4{ v^2}=\frac 1 {f^2}=\frac 2 {f_1^2}+\frac 1 {f_2^2}.
\label{v}
\ee

Let us now  include  a scalar field $h$, singlet
under the group $G\otimes H\otimes P$. For the  moment we will not be interested
in the self-couplings of this field $h$, and we will consider only
interaction terms with the vector fields  linear or quadratic in $h$.
The inclusion of a composite Higgs state was already considered for the general vector and axial-vector model 
in \cite{Casalbuoni:1997bv}. Here
we specialize it  in the context of the 4-site model.

The inclusion of a singlet $h$, by taking into account only dimension-four operators, is  straightforward:
\bea
{\cal L}_{hG}&=&
 (2 a_h \frac h v +b_h \frac {h^2}{v^2})
f_1^2[D_\mu \Sigma_1)^\dagger D^\mu \Sigma_1+(D_\mu \Sigma_3)^\dagger D^\mu \Sigma_3]\nn\\
&+& (2 c_h \frac h v +d_h \frac {h^2}{v^2}) f_2^2(D_\mu \Sigma_2)^\dagger D^\mu \Sigma_2.
\label{LhG}
\eea

In principle, one could also add dimension-five operators modifying the coupling of 
the singlet to a pair of gauge bosons and also Yukawa 
terms $c_f m_f/v\bar f f h$ which could modify the $h$ production and decay properties.
More generally one could introduce a singlet field $\rho_i$ for each chiral field $\Sigma_i$ as in \cite{Casalbuoni:1996wa}. 
We expect  the masses of the two heaviest singlets to be related to the scale of the new vector bosons while the scale of the lightest one to the Fermi scale. In our present analysis we however concentrate on the case of only one Higgs
state being present in the model spectrum.
\subsection{Parameter space}
\label{sec:gauge}

In the unitary gauge, the 4-site model predicts two new triplets of gauge 
bosons, which acquire mass through the same non-linear symmetry breaking 
mechanism giving mass to the SM gauge bosons. Let us denote with 
$W_{i\mu}^\pm$ and $Z_{i\mu}$ ($i = 1, 2$) the four charged and two neutral 
heavy resonances appearing as a consequence of the gauge group extension, and 
with $W^\pm_\mu$, $Z_\mu$ and $A_\mu$ the SM gauge bosons. Owing to its gauge 
structure, the 4-site
model a priori contains seven free parameters: the $SU(2)_L\otimes U(1)_Y$ 
gauge couplings, $\tilde g$ and $\tilde g'$, the extra $SU(2)_{1,2}$ gauge 
couplings that we assume to be equal, $g_2= g_1$, due to the $P$ symmetry, the bare 
masses of lighter ($W_1^\pm, Z_1$) and heavier ($W_2^\pm, Z_2$) gauge boson 
triplets, $M_{1,2}$, and their bare direct couplings to SM fermions, $b_{1,2}$,
as described in \cite{Accomando:2008jh,Accomando:2008dm}. 
However, their number can be reduced to four, by fixing the gauge couplings 
$\tilde g,\tilde g', g_1$ in terms of the three SM input parameters 
$e, G_F, M_Z$, which denote electric charge, Fermi constant and $Z$ boson mass, 
respectively. As a result, the parameter space is completely defined by  
four independent free parameters, which one can choose to be: $M_1$, $z$, $b_1$ and 
$b_2$, where $z=M_1/M_2$ is the ratio between the bare masses. In terms of 
these four parameters, physical masses and couplings of the extra gauge bosons 
to ordinary matter can be obtained via a complete numerical algorithm. This is 
one of the main results of \cite{Accomando:2011vt}, where this computation
was described 
at length, so we refer the reader to it for further details.
The outcome is the ability to reliably and accurately describe the 
full parameter space of the 4-site model even in regions of low mass 
and high $z$ where previously used approximations would fail. In the 
following, we choose to describe the full parameter space via the physical 
observables: other than $z$ (which, as shown in 
\cite{Accomando:2011vt}, is a good approximation of the ratio between physical 
masses $M_{W_1}/M_{W_2}$ or $M_{Z_1}/M_{Z_2}$) we take  $M_{W_1}, a_{W_1}$ and $a_{W_2}$ which denote the mass 
of the lighter 
extra  gauge boson and the couplings of the lighter and heavier extra 
 gauge bosons to ordinary matter, respectively. 
 
In terms of the above quantities, the Lagrangian describing the 
interaction between gauge bosons and fermions has the following expression: 
\bea
\label{aw}
{\mathcal L}_{NC}&=&\bar{\psi} \gamma^\mu \left [- e \mathbf{Q}^f A_\mu
+ {a}_{Z}^f Z_\mu+{a}_{Z_1}^fZ_{1\mu}+ {a}_{Z_2}^fZ_{2\mu}  \right ]\psi,\nn\\
{\mathcal L}_{CC}&=&\bar{\psi} \gamma^\mu T^-   \left(
a_WW_{\mu}^+ +a_{W_1} W_{1\mu }^+ +a_{W_2} W_{2\mu }^+\right)\psi + {\rm {h.c.}}
\eea
for the neutral and charged gauge sector, respectively. In the above 
formulae, $\psi$ denotes generally SM quarks and leptons. These expressions will be
used later on, when discussing production and decay of the extra gauge 
bosons. 

Before performing any meaningful analysis, it is mandatory to evaluate the
ensuing theoretical and experimental constraints on the parameter space of the model,
which we are going to do in the next two subsections.

\subsection{Theoretical constraints}
\label{sec:th-bounds}

\begin{figure}[!htb]
\begin{center}
\unitlength1.0cm
\begin{picture}(7,4)
\put(-1,-4){\epsfig{file=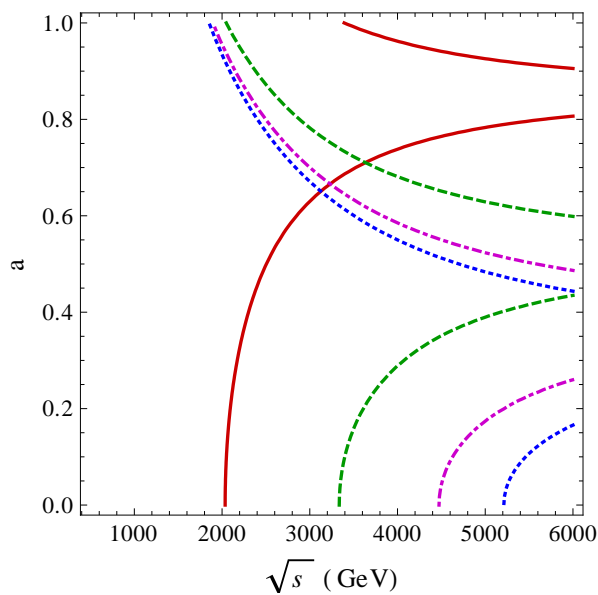,width=8cm}}
\end{picture}
\end{center}
\vskip 3.cm
\caption{Perturbative unitarity bounds: the allowed region is on the left side of the curves, we include both the vector and scalar contributions for four different values of $z$: 0.4 (purple-dashed-dotted line), 0.6 (blue-dotted line), 0.8 
(green-dashed line) and 0.95 (red-solid line).}
\label{fig:unitarity}
\end{figure}

One of the effects of including a scalar singlet in the 4-site model is a modification of the perturbative unitarity bounds 
acting in this scenario, which were derived in
\cite{Accomando:2008jh}, where the equivalence theorem was used in order to relate, at high energy,
the gauge boson scattering amplitudes to the corresponding Goldstone ones.
Using
\be
\Sigma_i=\exp{(i \frac {f}{2f_i^2}\vec \pi\cdot \vec \tau)},\,\,\,i=1,2,3,
\ee
where $\vec \pi$ are the Goldstones representing longitudinal $W$'s and $Z$'s, 
we get a coupling of the scalar $h$ boson to the $\vec \pi$ given by:
\be
2 a  \frac h v \frac 1 2 (\partial_\mu\vec \pi)^2
\ee
with
\be
a=a_h(1-z^2)+c_hz^2.
\ee
By following the analysis in \cite{Accomando:2008jh}, the $\pi\pi$ scattering amplitude,
 for $s\gg M^2_{1,2}$, gives a  term growing linearly with $s$ 
\bea
A(s,t,u)&\sim& \frac s {v^2} (1-\frac 3 4 b (1-z^4)^2-a^2)\nn\\
&=& \frac s {v^2} (1-\frac 3 4 (1-z^2)(1+z^2)^2-a^2)\nn\\
&=& \frac s {4v^2}(1-3z^2+3z^4+3z^6-4 a^2).
\eea
Herein, $z=f_1/ \sqrt{f_1^2+2 f_2^2}\sim M_1/M_2$, the EW scale  $v$ is given in (\ref{v}), and $b$ is given 
in (\ref{choice}). By considering the zero-isospin partial wave matrix element for all the amplitudes with SM longitudinal gauge bosons as external states 
 and imposing the unitarity bound $|a_0|<1$ for the maximum eigenvalue, we get  the result shown by the curves in Fig. \ref{fig:unitarity}. The maximum energy scale, up to which perturbative unitarity can be delayed, depends on $z$ and $a$, which is related to the coupling of the $h$ scalar particle to the longitudinal $W$'s ($a=1$ for a SM Higgs).
In Fig.~\ref{fig:unitarity}
in particular we show the limits for the four $z$ values chosen for our forthcoming phenomenological study. 

The 4-site model has in addition two vector-boson triplets with, potentially, bad behaving longitudinal scattering amplitudes, so one has to require a fully perturbative regime for all involved particles. The unitarity limit must thus be extended, in order to ensure a good high energy behaviour for all scattering amplitudes, i.e., with both SM and extra gauge bosons as external states. 
However, since  in the following analysis we are interested in a mass range for $M_{1,2}$ below 2 TeV, we are on the safe side concerning the unitarity bound limits.

\subsection{Experimental constraints}
\label{sec:exp-bounds}

Universal EW radiative corrections to the precision observables 
measured by LEP, SLC and Tevatron experiments can be efficiently quantified in 
terms of three parameters: $\epsilon_1, \epsilon_2$, and $\epsilon_3$ 
(or $S$, $T$, and $U$)
\cite{Peskin:1990zt,Peskin:1992sw,Altarelli:1991zd,Altarelli:1998et}. 
Besides these SM contributions, the $\epsilon_i$ ($i=1,2,3$) parameters also allow one to 
describe the low-energy effects of potential heavy-mass new physics. For that 
reason, they are a powerful method to constrain theories beyond the SM.   Besides the indirect effects, in this section we will also derive bounds from direct searches of the extra gauge bosons at Tevatron and LHC and from the new measurements at the LHC of the decay rates of the Higgs boson. Lets start with the latter.

\subsubsection{Constraints from Higgs sector measurements}
\begin{figure}[t!]
\begin{center}
\vspace{-.8cm}
\unitlength1.0cm
\begin{picture}(7,10)
\put(-4.3,2.7){\epsfig{file=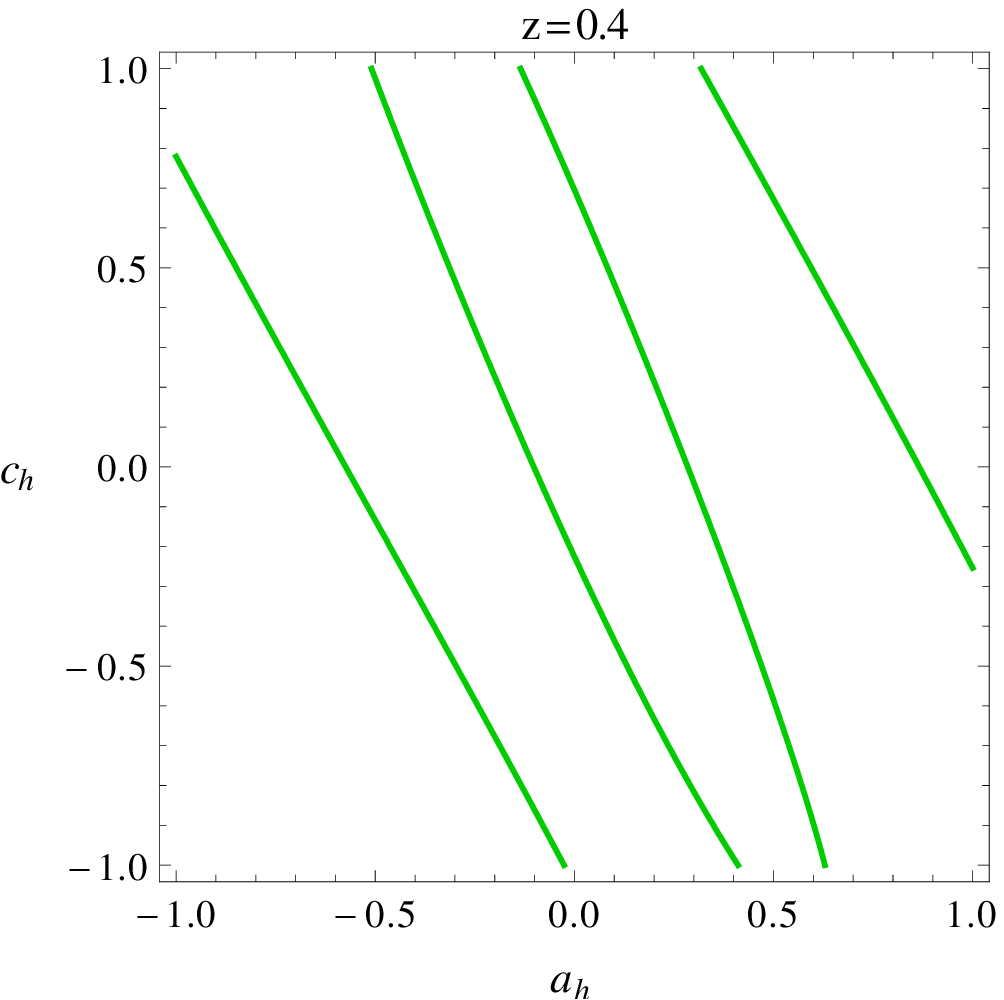,width=7cm}}
\put(3.5,2.7){\epsfig{file=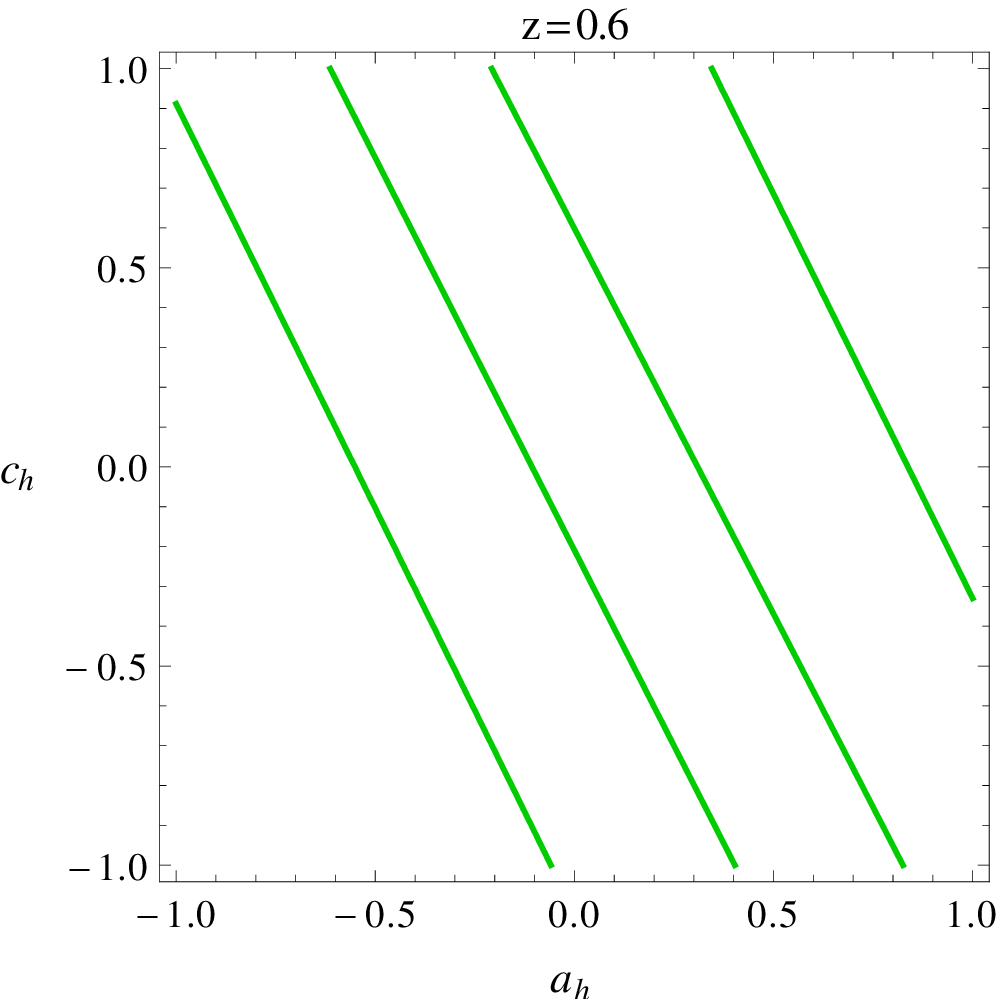,width=7cm}}
\put(-4.3,-5){\epsfig{file=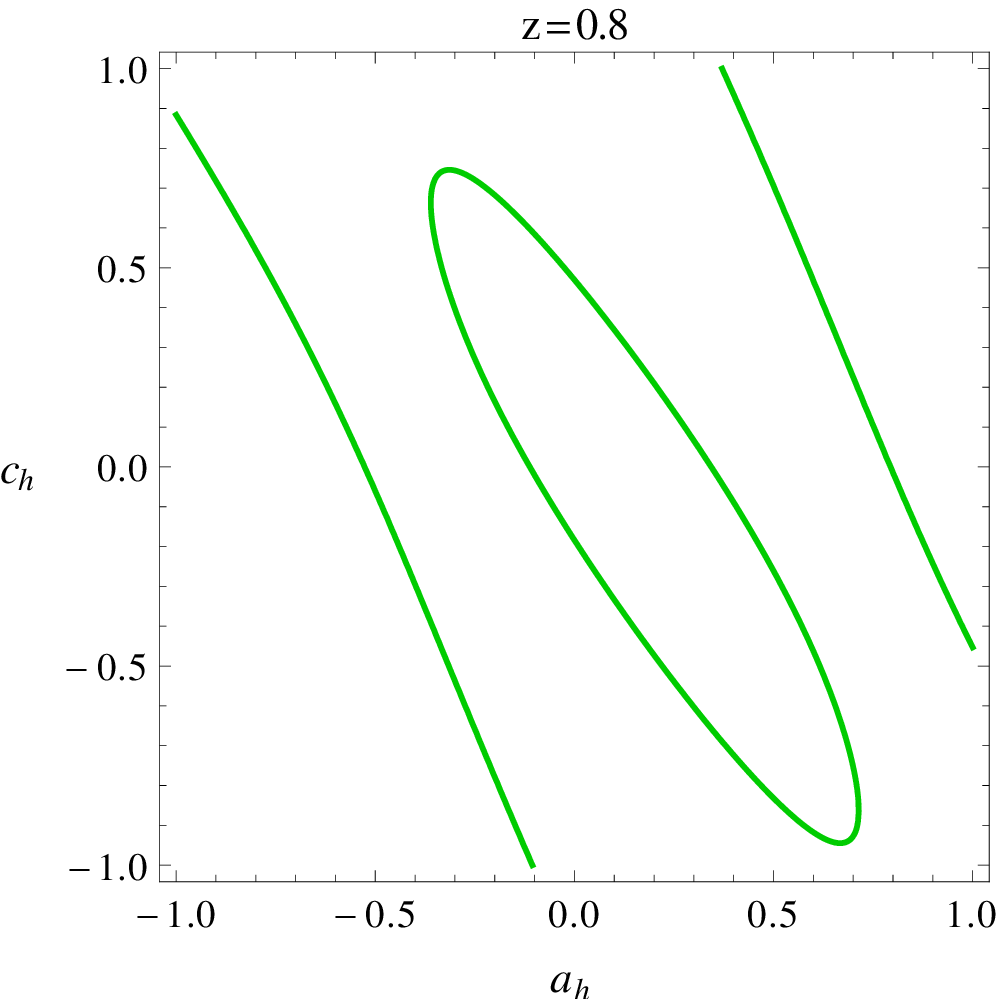,width=7cm}}
\put(3.5,-5){\epsfig{file=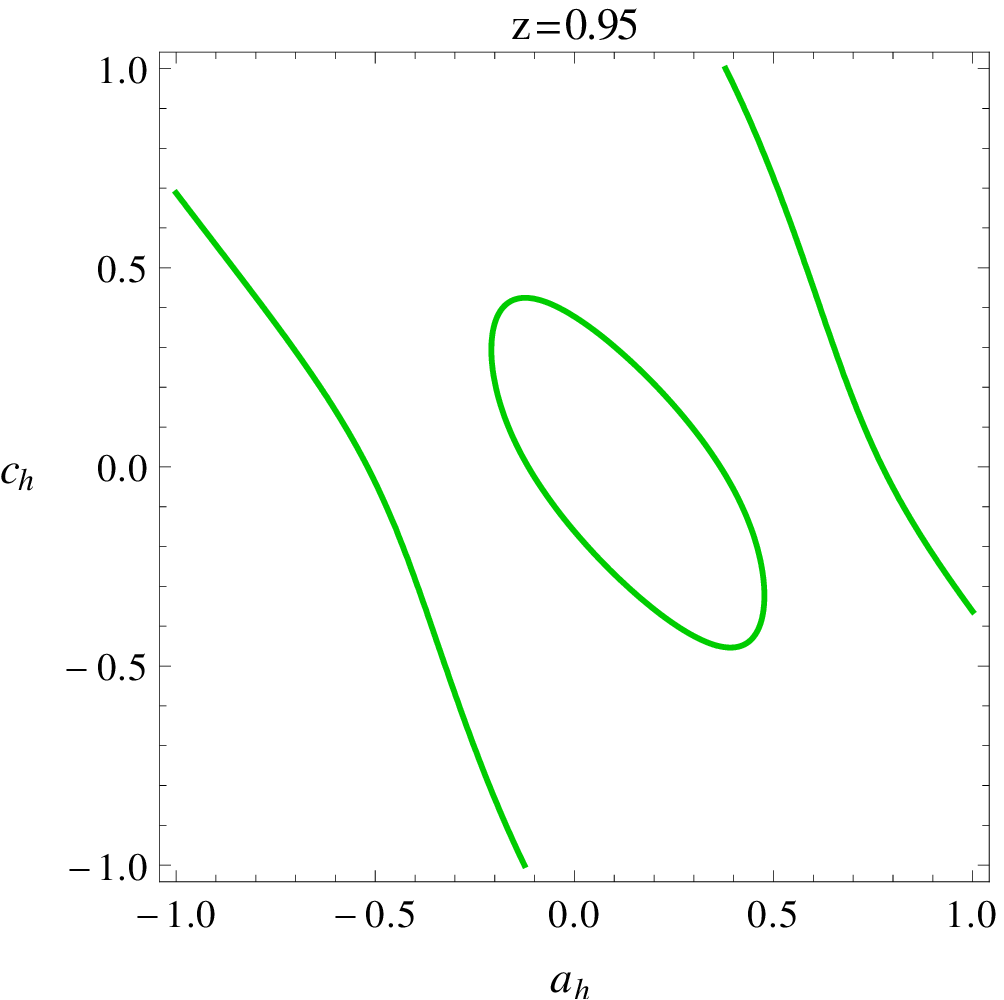,width=7cm}}
\end{picture}
\end{center}
\vskip 4.cm \caption{
95\% Confidence Level
(CL) limits in the plane ($a_h,c_h$), the allowed region
is between the two external lines and outside the central hole or central lines (as applicable).
The $z$ parameter is fixed to be $z=0.4$, 0.6, 0.8, 0.95. We assume $c_f=1$.
We have used the Higgs boson rates from CMS  for the $\gamma\gamma$, $WW$ and $ZZ$ channels, as estimated in \cite{lpccgunion,Gunion:2012gc}. Similar figures are obtained with ATLAS results.}
\label{fig:ahch}
\end{figure}

As we have noticed, the composite Higgs sector can be parametrized using $z$, $a_h$ and $c_h$:  these parameters are bounded from recent measurements performed
at the LHC \cite{incandela,gianotti}. In our analysis, which is very preliminary, just like these LHC data are,  we have used  the results extracted from the rates of the processes  $H\rightarrow\gamma\gamma,ZZ,WW$ 
\cite{lpccgunion,Gunion:2012gc}, to get bounds on the parameter plane ($a_h,c_h$). 
In our model,
the loop contribution to the di-photon decay mode of the Higgs boson has additional components from the loops of 
$W_1$ and $W_2$. Therefore, we have to re-evaluate the rate for $pp\to h\to \gamma\gamma$
in presence of the latter and compare its value against experimental limits, while at the same time
ensure that the rates for $pp\to h\to ZZ$ and $WW$ also remain consistent with experiment. We list here the couplings of the singlet $h$ state to the charged
gauge bosons of our model:
\bea
&&\frac {2h}{v} \large [a M^2_W W^+ W^- + a_h M_1^2 W^+_1W^-_1\nn\\
&&+ (a_h z^2+c_h (1-z^2)M_2^2 W^+_2W^-_2\large].
\eea
The results are summarised in Fig.~\ref{fig:ahch}  for the four chosen values of $z$. 
Besides these bounds,
one has also to take into account that contraints on 
the plane $(a,c_f)$  are already  available, so that  $a$ cannot be very different from 1, depending on 
$c_f$ \cite{Azatov:2012ga,Ellis:2012hz,Espinosa:2012im,petrucciani,Giardino:2012dp} (in our present analysis we assume $c_f=1$). 

Moreover,
if  $a\neq 1$ one has to add  additional  model contributions to the  $S$ and $T$ parameters.
The contributions  from a non-standard scalar sector can be summarised through  additional terms entering
the expression for $\epsilon_1$ and $\epsilon_3$:
\be
\Delta\epsilon_1^h=-\frac {3\alpha} {16\pi c^2_{\theta_W}} (1-a^2)\log\left(\frac{\Lambda}{M_1}\right)\quad 
\Delta\epsilon_3^h=-\frac \alpha {48\pi s^2_{\theta_W}} (1-a^2)\log\left(\frac{\Lambda}{M_1}\right).
\label{Deps}
\ee
 In order to minimise these extra contributes to the $S$ and $T$ parameters, we will choose values of $a_h$ and $c_h$, inside the allowed regions of Fig. \ref{fig:ahch}, which give  $a$ as much close to 1 as possible. In Tab.~\ref{tab:ahch}
we summarise the values for $a_h,c_h$ and $ a$ that  we will use for our upcoming phenomenological analysis, obtained by minimising the $\Delta\chi^2$ built  from the experimental bounds on $H\rightarrow\gamma\gamma,ZZ,WW$.
\begin{table}[!htb]
\begin{center}
\begin{tabular}{||c|c|c|c||}
\hline \hline
$z$&$a_h$ & $c_h$& $a$ \\
\hline 
\hline
0.4&1.00&$-0.49$&0.76\\
\hline
0.6&0.25&1.00&0.52\\
\hline
0.8& 0.26& 1.00& 0.73\\
\hline
0.95& 0.19& 1.00& 0.92\\
\hline\hline
\end{tabular}
\end{center}
\caption{Values for $a_h$ and $c_h$ chosen in order to maximise $a$ for each $z$. }
\label{tab:ahch}
\end{table}
It is clear that, for $z=0.6$, the non-standard scalar contribution is quite large, instead for $z=0.95$ it is very marginal.

\subsubsection{Constraints from EWPTs}
In \cite{Accomando:2011vt}, 
a complete 
numerical calculation of all $\epsilon_i$ ($i=1,2,3$) parameters at tree level in the 4-site model, 
going beyond popular approximations used in the past, was carried out and 
a combined fit to the experimental results taking into account their full 
correlation, extracted. The exact 
results allowed one to span the full 
parameter space of the model, reliably computing  regions characterised by 
small $g_1$ (or $M_1$) values and sizeable $b_{1,2}$ bare couplings.

This analysis can be straightforwardly applied to the model at hand with a singlet scalar $h$ included,
by adding the corresponding contributions in (\ref{Deps}), with the numerical choices of Tab. \ref{tab:ahch} and $\Lambda=3$ TeV, to the SM values evaluated for $m_H=125$ GeV.

\begin{figure}[!t]
\begin{center}
\unitlength1.0cm
\begin{picture}(7,4)
\put(-5.6,-4){\epsfig{file=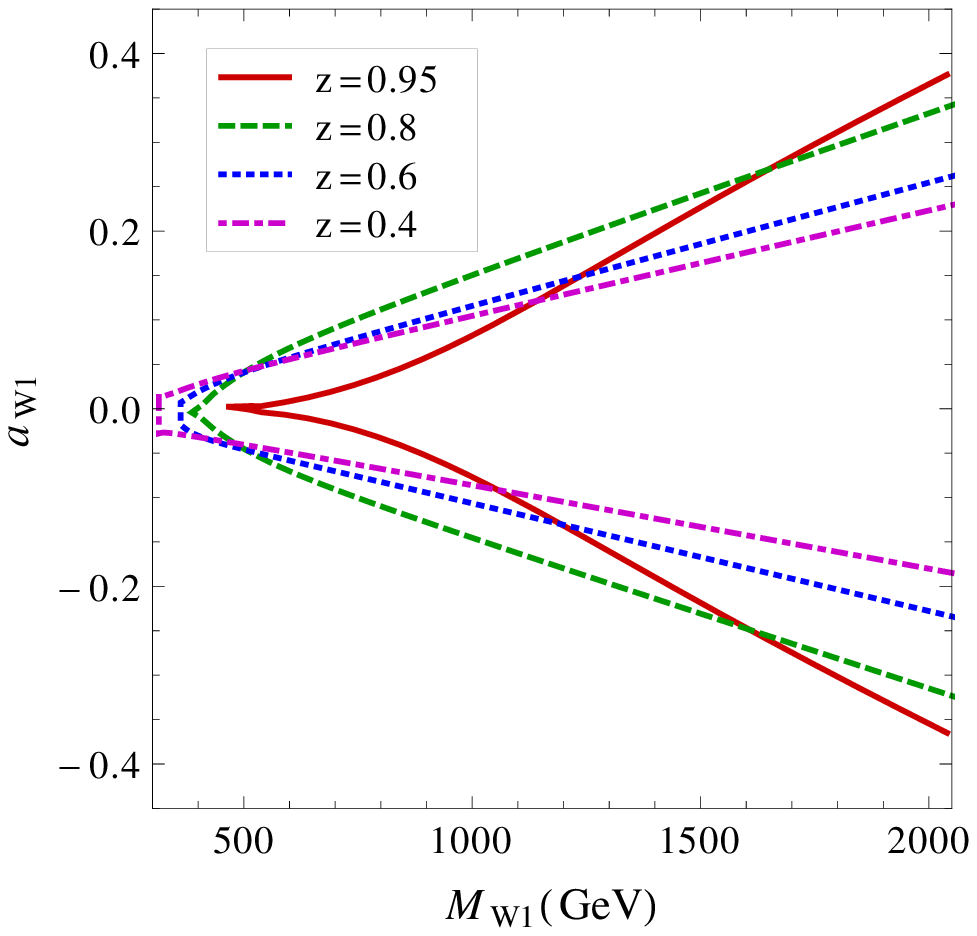,width=7.5cm}}
\put(3.5,-4){\epsfig{file=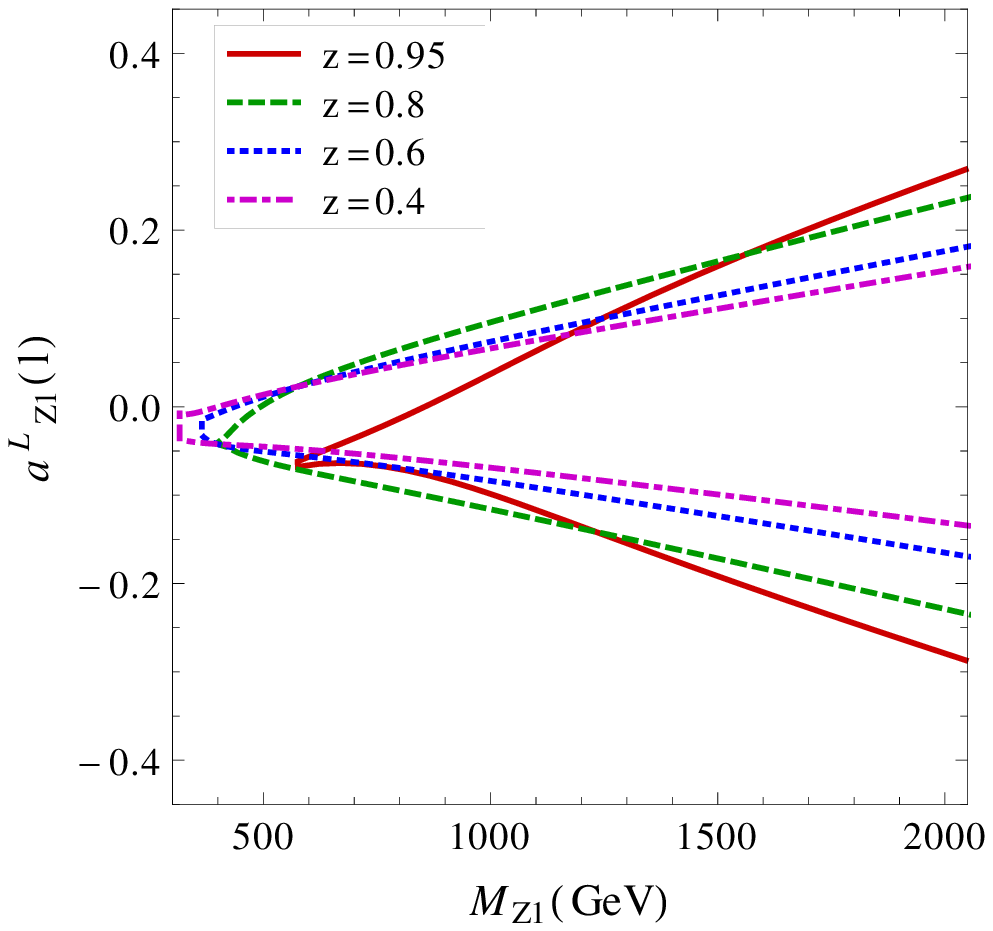,width=7.5cm}}
\end{picture}
\end{center}
\vskip 3.5cm
\caption{Left: 95$\%$ CL EWPT bounds in the parameter space given in 
terms of physical mass, $M_{W_1}$,  and coupling between the lighter 
extra charged gauge boson and SM fermions, $a_{W_1}$. Right: 
the same in the neutral plane
($M_{Z_1}$, $a_{Z_1}^L(l)$), where $a_{Z_1}^L(l)$ represents the coupling of the  left-handed charged lepton to the $Z_1$ boson.
We consider four reference $z$ values: $z= 0.4$, 0.6, 0.8 and 0.95. The 
allowed regions are delimited by the curves.}
\label{fig:EWPT_a1-a2}
\end{figure}

In Fig.~\ref{fig:EWPT_a1-a2} we show the limits on the 4-site model 
supplemented by one active composite Higgs scalar with mass at 125 GeV, 
in both the charged and neutral plane, over which we define CL
regions according to Gaussian statistics.
Due to the fact that EWPTs impose a stringent correlation between $a_{W_1}$ and $a_{W_2}$, the number of free parameters can 
further be reduced to three by  choosing $a_{W_2}$ to be  maximal, in 
absolute value, once $M_{W_1}$, $a_{W_1}$ and $z$ are fixed.
From Fig.~\ref{fig:EWPT_a1-a2} we deduce  that, even if constrained, the $a_{W_1}$ coupling can be of the same order of 
magnitude as the corresponding SM coupling. This result is common to all
other couplings between extra gauge bosons and ordinary matter, which can
uniquely be derived from $a_{W_1}$ via the aforementioned numerical algorithm. 
An additional information that one can extract from
Fig.~\ref{fig:EWPT_a1-a2} concerns the minimum mass of the extra gauge bosons allowed by EWPTs. As one can see, its value 
depends on the $z$ parameter and can range between 300 and 500 GeV. In the right panel of Fig.~\ref{fig:EWPT_a1-a2} the same bounds are shown in the plane
($M_{Z_1}$, $a_{Z_1}^L(l)$).

\subsubsection{Constraints from gauge sector measurements}
In the remainder of this section we perform a brief review of the experimental bounds on the 4-site model coming
from direct searches of $W'$ and $Z'$ bosons via DY channels into leptons. Clearly, in the latter (and limited to the neutral DY process), there cannot be any perceptible contribution due to the additional Higgs scalar present in our model, as the latter couples negligibly to both the initial state quarks and the final state leptons. 

\begin{figure}[!t]
\begin{center}
\vspace{-.8cm}
\unitlength1.0cm
\begin{picture}(7,10)
\put(-4.3,2.7){\epsfig{file=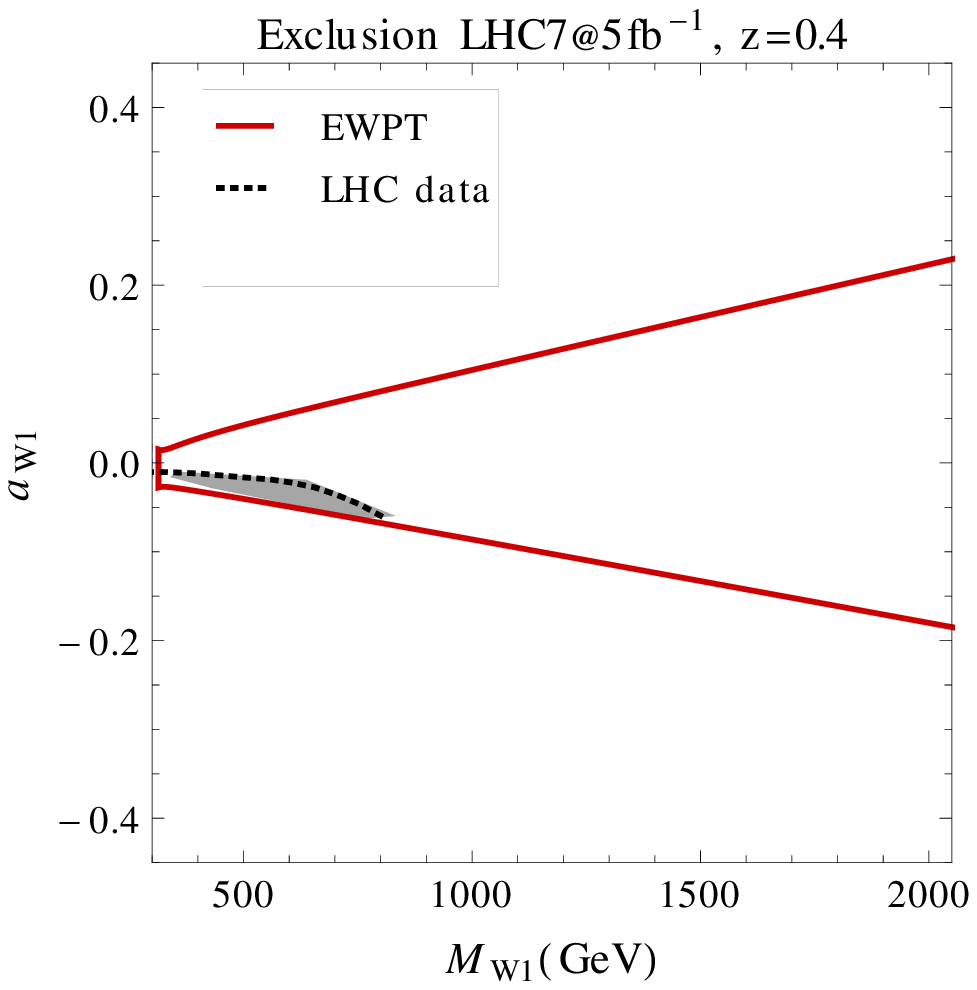,width=7.5cm}}
\put(3.5,2.7){\epsfig{file=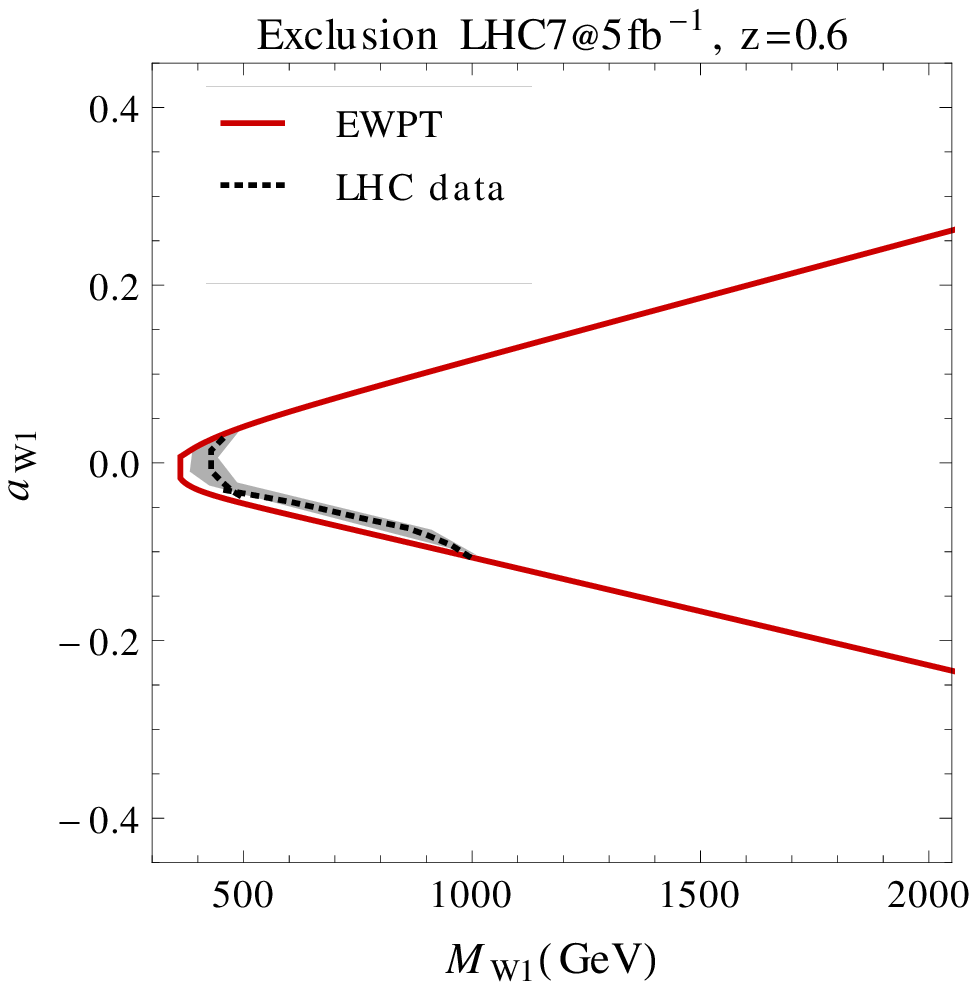,width=7.5cm}}
\put(-4.3,-5){\epsfig{file=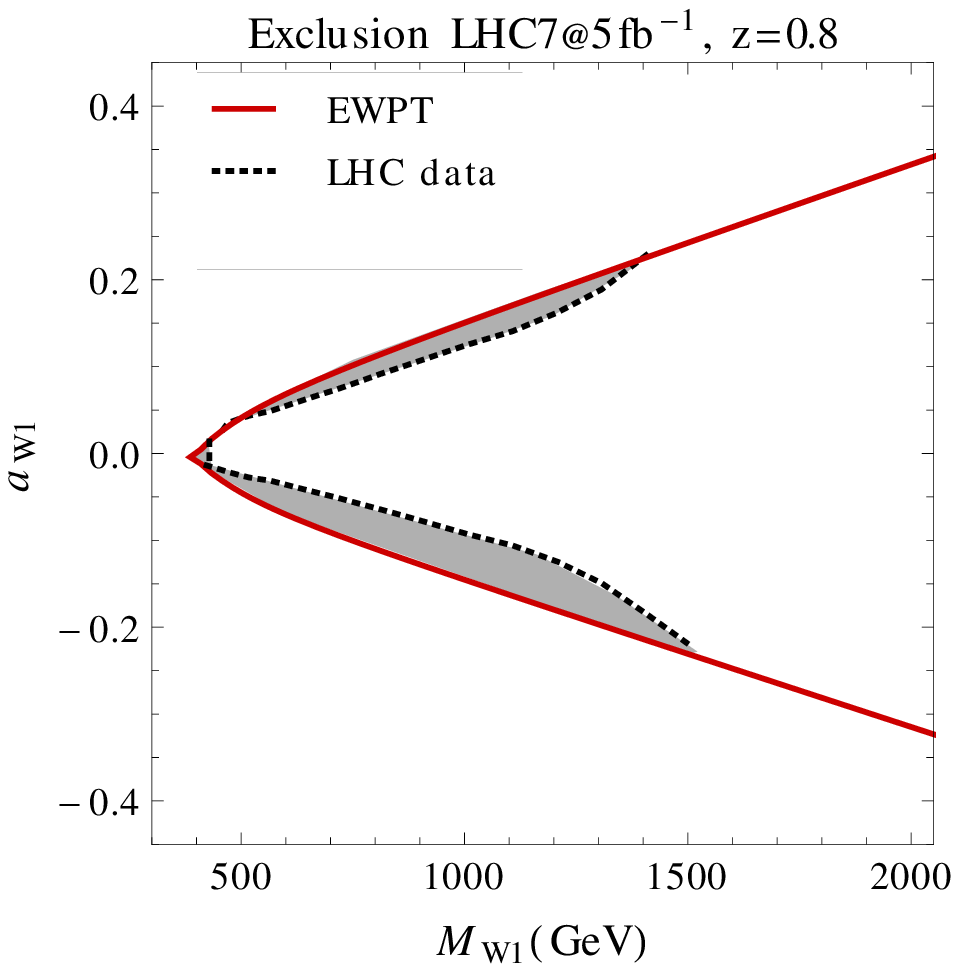,width=7.5cm}}
\put(3.5,-5){\epsfig{file=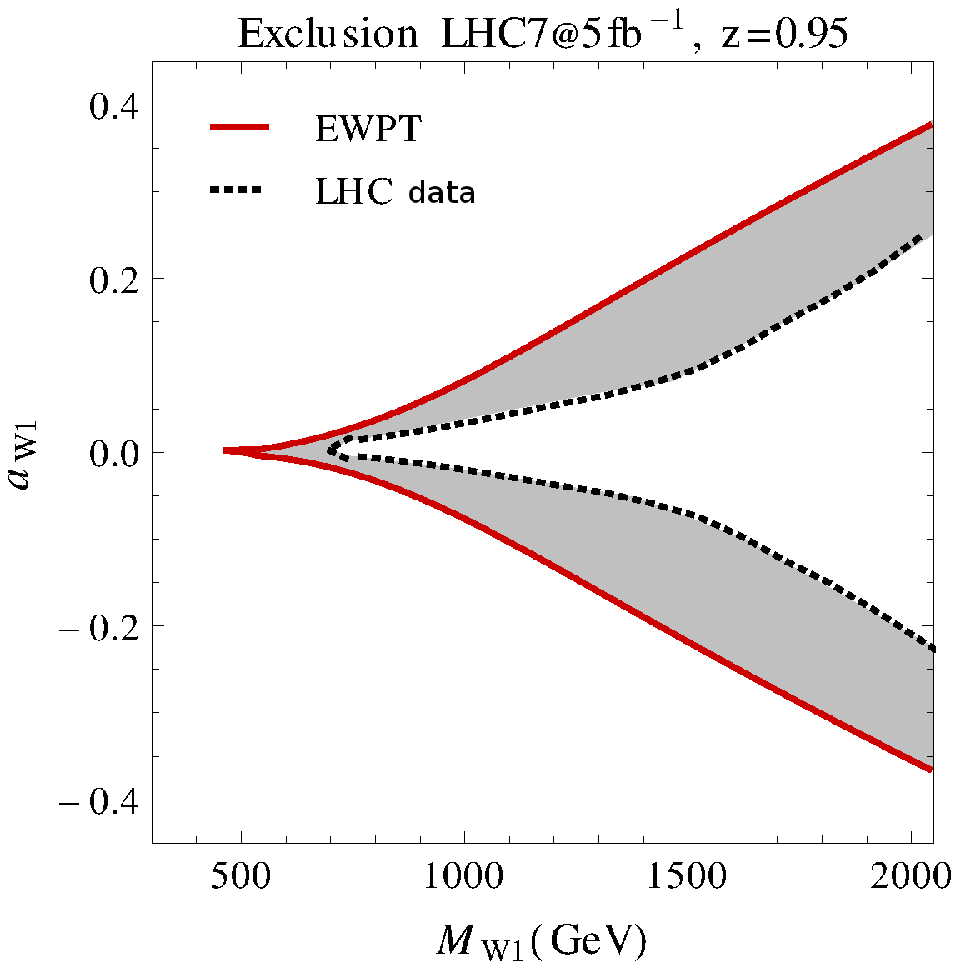,width=7.5cm}}
\end{picture}
\end{center}
\vskip 4.cm \caption{
95\% CL exclusion limits in the $(M_{W_1},a_{W_1})$ plane at the 7 TeV LHC with an integrated luminosity of 5 fb$^{-1}$ 
considering the direct limits on the cross section (black-dotted line).
The red-solid contour defines the parameter space allowed by EWPTs. 
The $z$ parameter is fixed to be $z=0.4$, 0.6, 0.8, 0.95. 
}
\label{fig:W1_E_5fb}
\end{figure}

We have considered  the last published results from ATLAS and
CMS at LHC at 7~TeV and 5~fb$^{-1}$~\cite{ATLAS-CONF-2012-007,Chatrchyan:2012qk} and  extracted both limits from $Z'$ and $W'$ searches.
We find that those from $W'$ are more stringent with respect to those from $Z'$ searches. So, in the following
we will consider only the limits stemming from 
searches for charged particles.  They are shown in Fig.~\ref{fig:W1_E_5fb}  in the plane
($M_{W_1}$,$a_{W_1}$) for the four reference $z$ values (0.4, 0.6, 0.8, 0.95).
These bounds are obtained by mapping  the limits from direct searches onto limits on the 
cross section.
%

So far we have reported experimental limits from DY direct searches based on currently available data. However, the ultimate goal
of our analysis is to compare the scope of the LHC at all its energy and luminosity stages in accessing the parameter space
of our model in either DY processes or the charged di-boson mode, for which there are currently no direct limits (on the cross section or else).
So, we are bound in the remainder of the paper, in order to compare their relative yield, to use simulated data. Clearly, to be confident
that we are accurately repeating the salient features of a proper experimental analysis, we must compare the limits that we obtain by using
simulated data with those extracted from the real ones. We can of course do so only in the case of the DY modes. 
 
\begin{figure}[h!]
\begin{center}
\vspace{-.8cm}
\unitlength1.0cm
\begin{picture}(7,10)
\put(-4.3,2.7){\epsfig{file=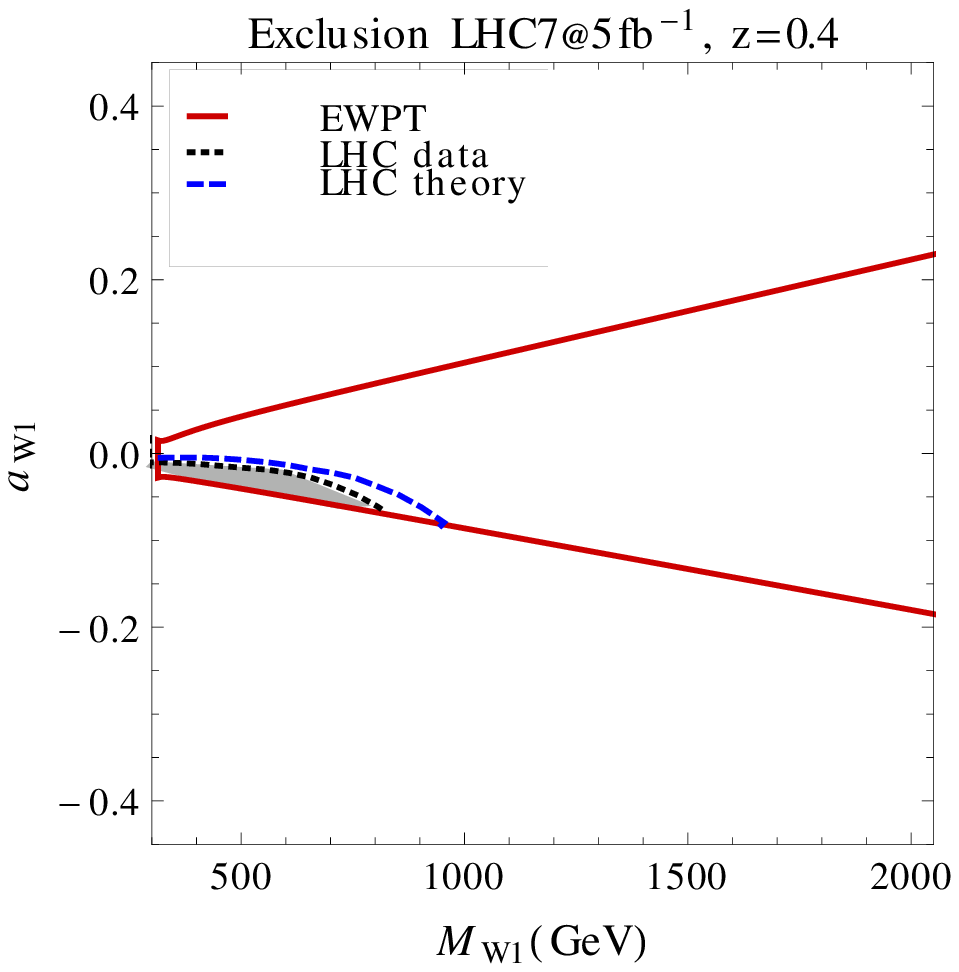,width=7.5cm}}
\put(3.5,2.7){\epsfig{file=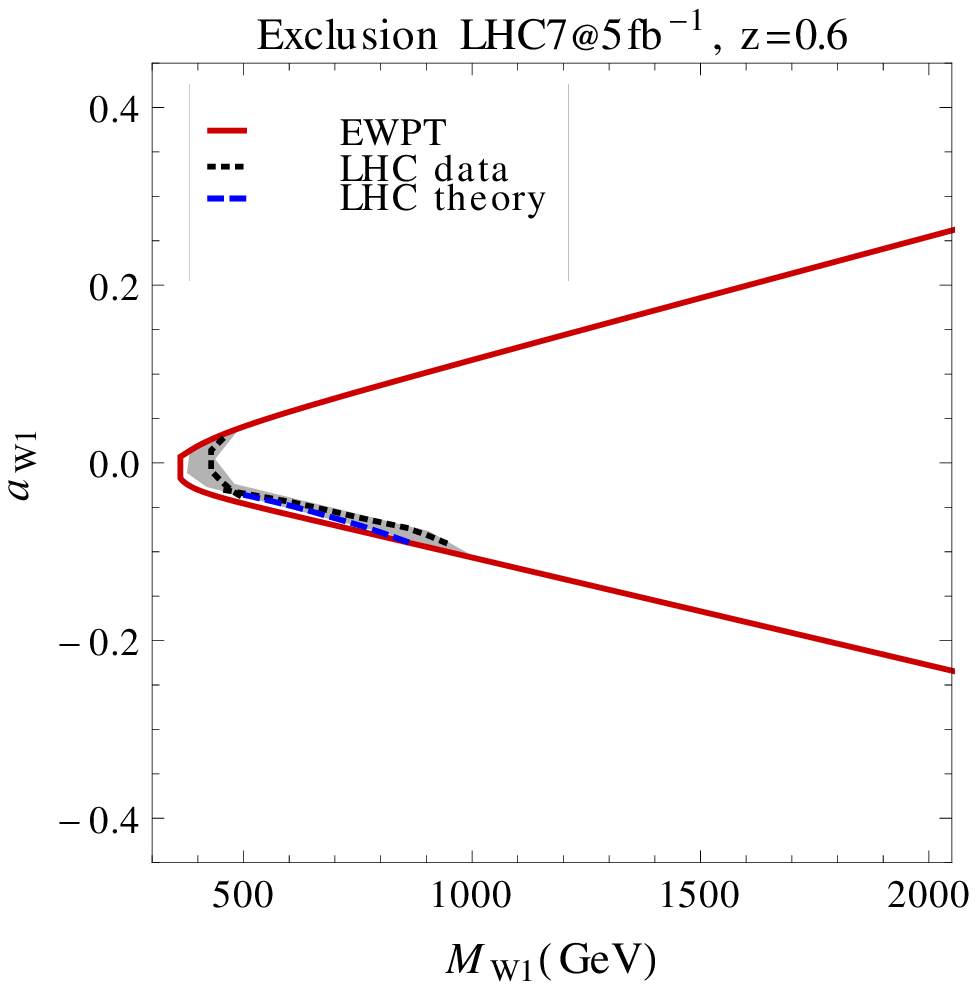,width=7.5cm}}
\put(-4.3,-5){\epsfig{file=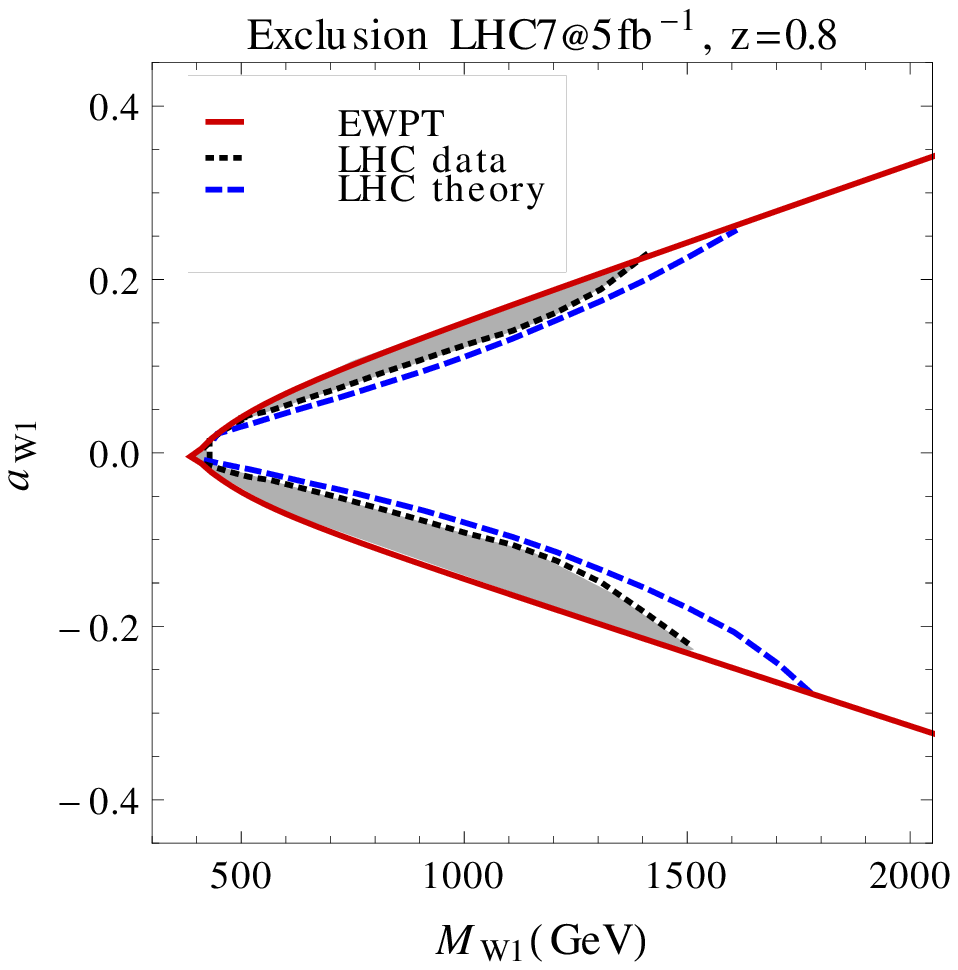,width=7.5cm}}
\put(3.5,-5){\epsfig{file=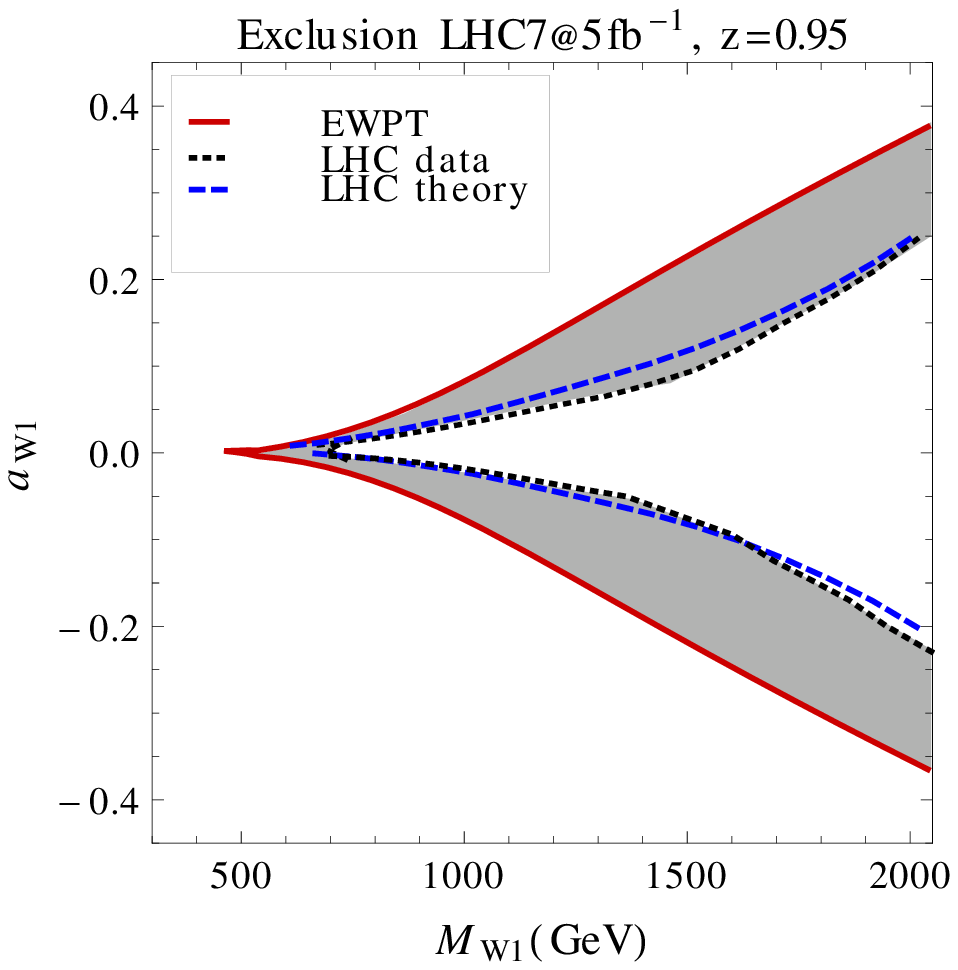,width=7.5cm}}
\end{picture}
\end{center}
\vskip 4.cm \caption{
95\% CL exclusion limits in the $(M_{W_1},a_{W_1})$ plane at the 7 TeV LHC with an integrated luminosity of 5 fb$^{-1}$ 
considering the direct limits on the cross section and the theoretical ones, performed as described in the text.
The red-solid contour defines the parameter space allowed by EWPTs. 
The $z$ parameter is fixed to be $z=$0.4, 0.6, 0.8, 0.95. 
}
\label{fig:W1_E_exp+th_5fb}
\end{figure}

We proceed then as follows. Taking exclusion, for example, we consider the bounds on the parameter space requiring a statistical significance 
lower than 2, which means:
\be\label{eq:signi}
S=\frac{T-B}{\sqrt{\mathcal{B}}}<2\quad\mbox{with}\quad \mathcal{B}=\left\{\begin{array}{c}
                                                                            B\quad\mbox{if}\quad B\ge1\\
                                                                            1\quad\mbox{if}\quad B< 1
                                                                           \end{array}\right.
\ee
where $T$ and $B$ are, respectively, the total and the expected (from background) numbers of events. However, applying this method to simulated data gives 
 different
 results from those obtained by the experiments (we are assuming the same acceptance and selection cuts, albeit at the
parton level), 
in particular, the theoretical approach gives 
more stringent bounds. This is due to the fact that we consider the full cross section, without
including any kind of experimental efficiency to detect the final state over the volume defined by our cuts, so that
 our number of events is higher than 
the experimental one, and so in turn the cross section and the statistical significances are larger. We note however that, if we consider an efficiency between 
$50\%$ 
and 30$\%$, decreasing with the mass of the resonance entering the DY mode, then we reproduce quite well the experimental bounds.
In Fig.~\ref{fig:W1_E_exp+th_5fb}  we show the two different limits, including also the mentioned efficiency for the theoretical ones.
The consistency between the two is excellent. Therefore, we feel confident that to adopt these efficiency measures will enable us to reproduce 
accurately the 
eventual experimental findings
assuming data sets that are not currently available. We will proceed in the same way for the case of di-boson events as well, after all the efficiency 
values
above are essentially extracted as an average between rates applicable to pairs of electron and muon separately (from DY), whereas for di-boson
events we are looking at one electron-muon pair. Only addition that we ought to account for in the latter case is to
estimate the efficiency to detect the missing transverse energy, who does not enter in the former case. We estimate this to be 70\% and mass 
independent \cite{Chatrchyan:2011tn}. 
  
\section{Di-boson production and decay}

We describe in this section the phenomenology of process (\ref{eq:process}), hereafter sometimes referred to for simplicity as
$e\mu +2\nu$ production, from the point of view of both its production and decay dynamics.

\subsection{Decay phenomenology}

Here we summarise the decay properties, i.e., widths and Branching Ratios (BRs), of the heavy gauge bosons, $W_{1,2},Z_{1,2}$, 
predicted by the 4-site model. A first peculiarity of the 4-site model is related to 
the nature of the two triplets of extra gauge bosons and their mass hierarchy (the gauge bosons of the same triplet are 
almost degenerate in mass, which means $M_{Z_{1,2}}\simeq M_{W_{1,2}}\simeq M_{1,2}$). 
The lighter triplet, $W^\pm_{1},Z_{1}$, are vector 
bosons while the heavier ones, $W^\pm_{2},Z_{2}$, are axial-vectors (neglecting EW corrections). 
Unlike closely related
models, like walking technicolor~\cite{Belyaev:2008yj}, no mass spectrum inversion is possible. The mass splitting, 
$\Delta M\simeq M_{W_2}-M_{W_1}\simeq  M_{Z_2}-M_{Z_1}\simeq M_2-M_1$, is always positive and its size depends on the free 
$z$ parameter. Here is an approximate relation, which works though for $M_1>400\mbox{~GeV and~}z<0.9$: 
\be
\Delta M\simeq\frac{1-z}{z}M_{W_1}\simeq\frac{1-z}{z}M_{Z_1}\simeq\frac{1-z}{z}M_{1}, \qquad z>0.
\label{eq:DeltaM}
\ee  
The above eq.~(\ref{eq:DeltaM}) contains also information on the kind of multi-resonance spectrum we might
expect. Owing to the $z$ parameter dependence, there is no fixed relation between the two charged or neutral
gauge boson masses. We can thus have scenarios where the two pairs of resonances, $W^\pm_{1},Z_{1}$ and 
$W^\pm_{2},Z_{2}$, lie quite apart from each other, and portions of the parameter space in which they are (almost) degenerate. In the 
latter case, the multi-resonance signature distinctive of the 4-site model would collapse 
into the more general single ${W',Z'}$ signal. The 4-site model would thus manifest a degeneracy with well known 
scenarios predicting only two additional (one charged and one neutral) gauge bosons.

\begin{figure}[!t]
\begin{center}
\unitlength1.0cm
\begin{picture}(7,4)
\put(-5.6,-4){\epsfig{file=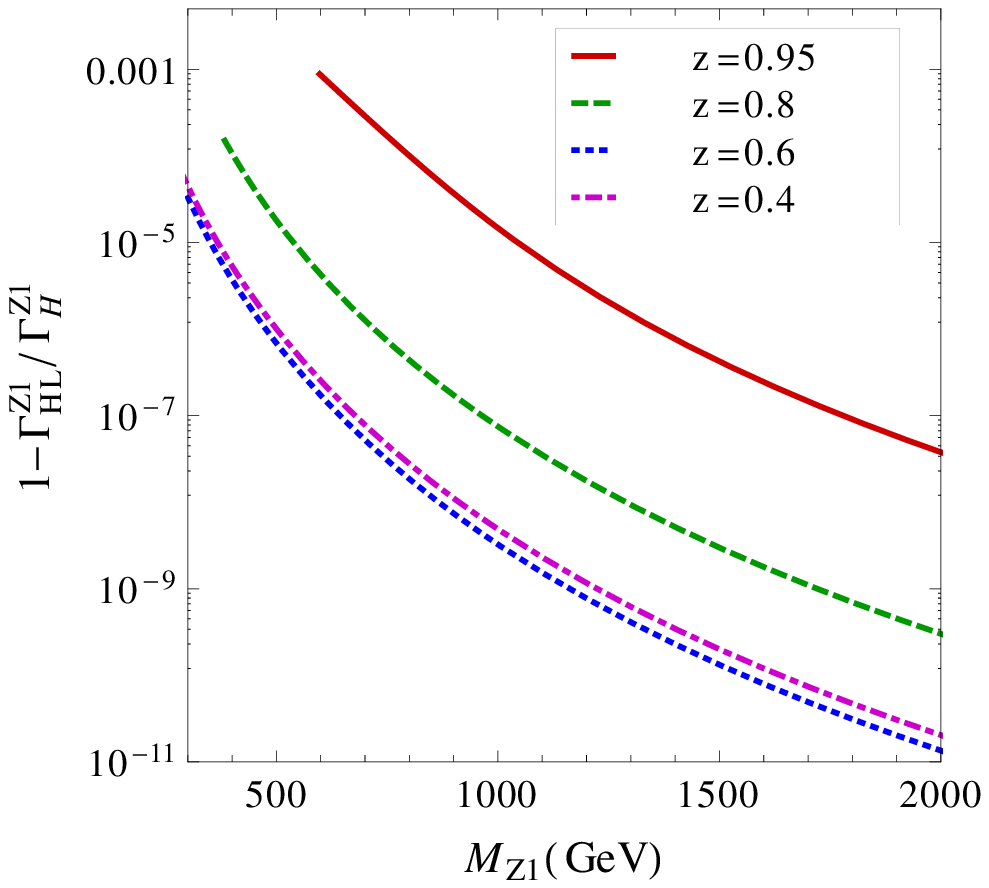,width=7.5cm}}
\put(3.5,-4){\epsfig{file=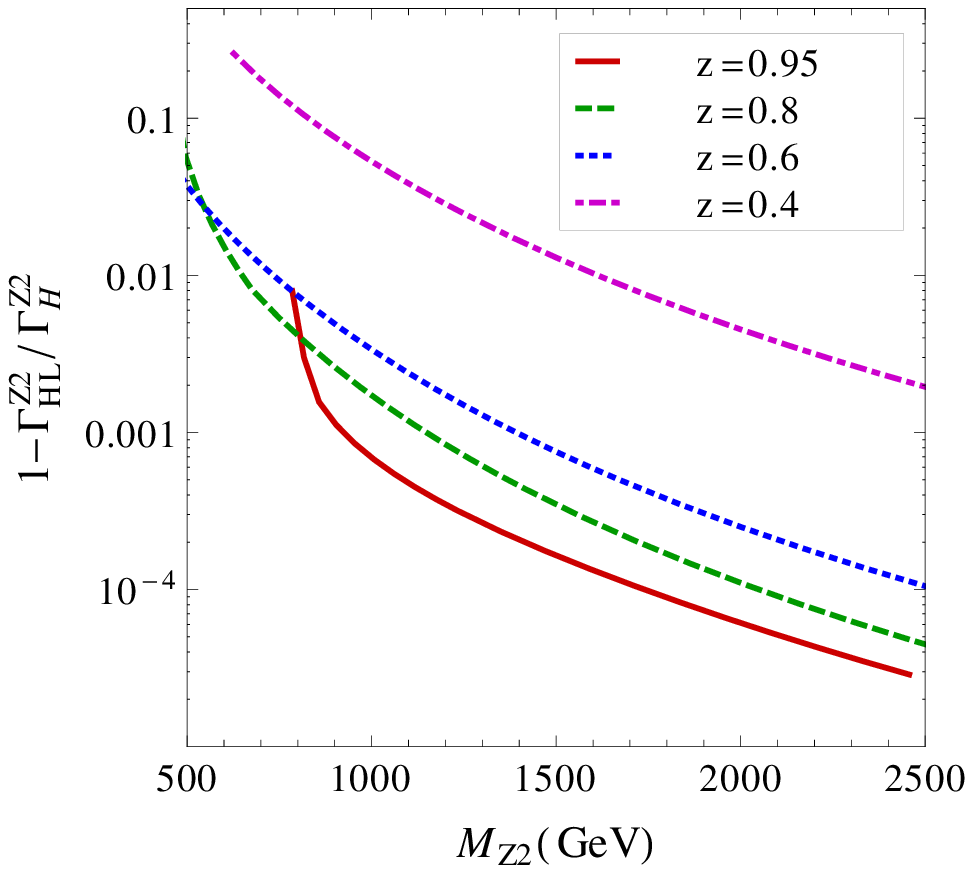,width=7.5cm}}
\end{picture}
\end{center}
\vskip2.5cm
\begin{center}
\unitlength1.0cm
\begin{picture}(7,4)
\put(-5.6,-4){\epsfig{file=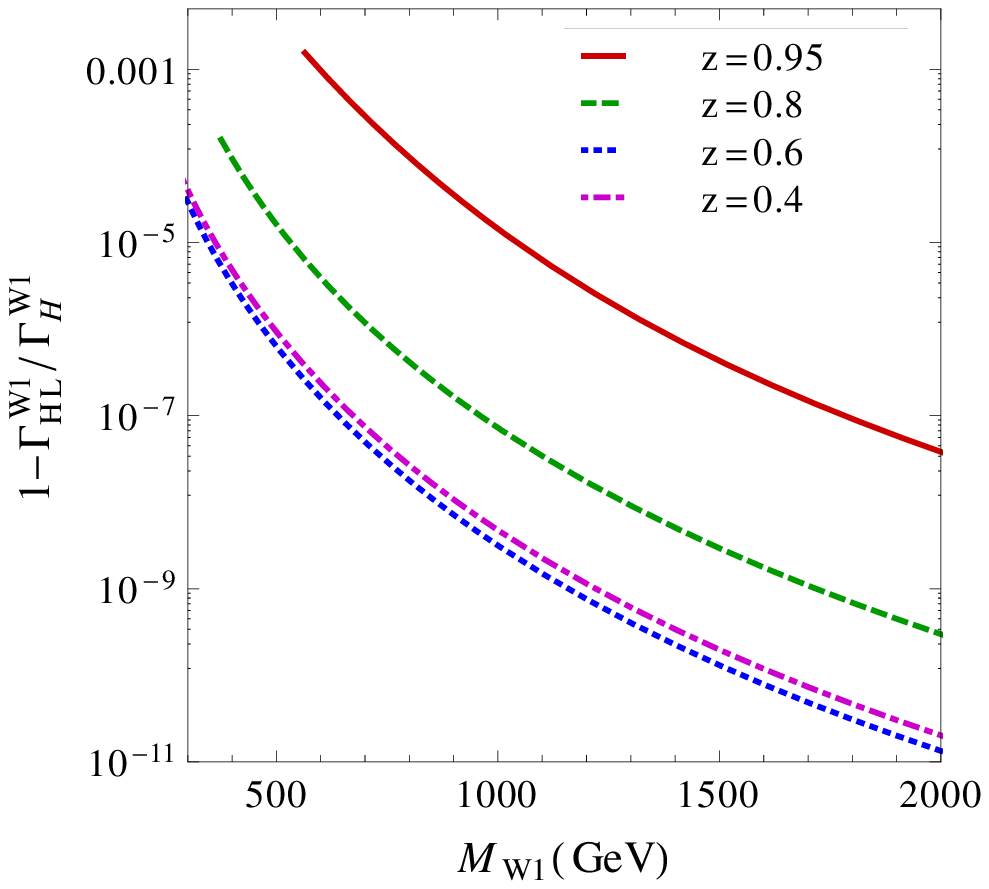,width=7.5cm}}
\put(3.5,-4){\epsfig{file=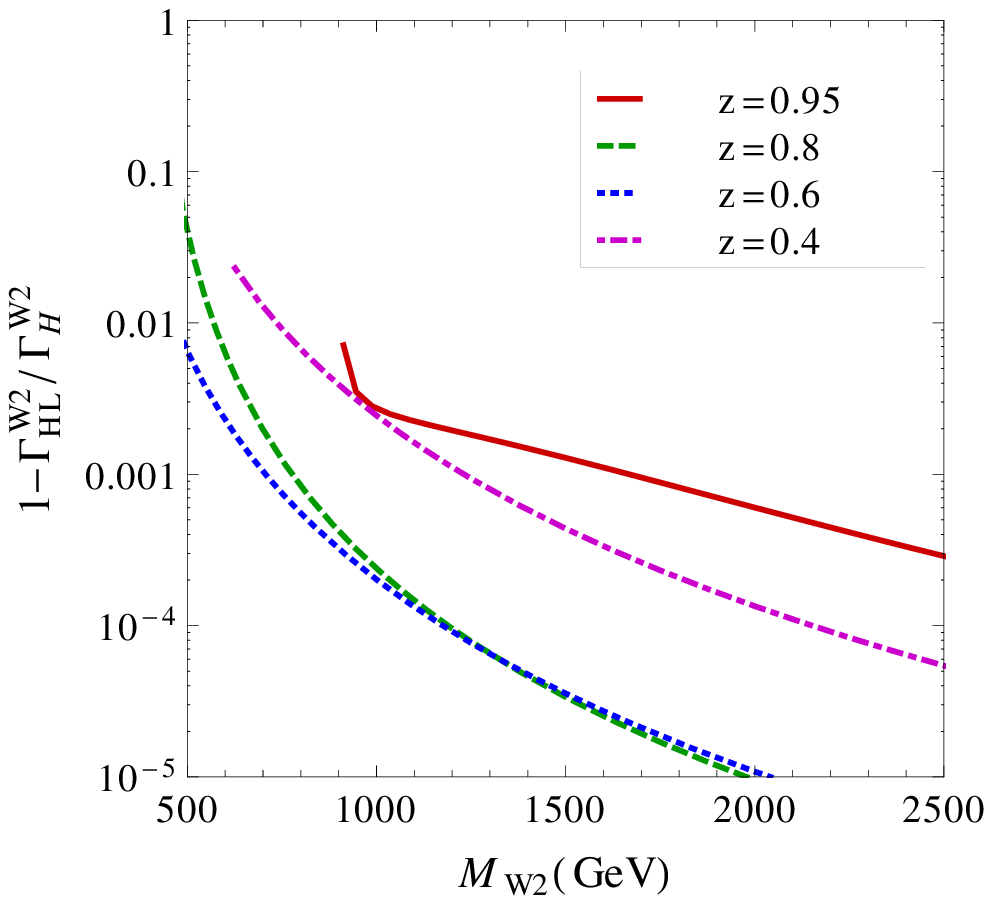,width=7.5cm}}
\end{picture}
\end{center}
\vskip 3.5cm
\caption{Top-left: Correction to the Higgsless decay rates as deviation from 1 in terms of the
ratio between the width of  the $Z_1$ decaying in everything except Higgs bosons ($\Gamma^{Z_1}_{HL}=\Gamma^{Z_1}_{H}-\Gamma^{Z_1}_{\rm Only~
Higgs}$) and the total width
 ($\Gamma^{Z_1}_{H}$). Top-right:  
the same for $Z_2$.  Bottom-left: the same for $W_1$. Bottom-right: the same for $W_2$.
The four usual choices for $z$ have been adopted.}
\label{fig:gammas}
\end{figure}

The widths and BRs of our four additional gauge states have been studied in previous papers by some of
us, see Refs.~\cite{Accomando:2011eu,Accomando:2011xi,Accomando:2010ir,Accomando:2008jh,Accomando:2008dm}. Those results were however relevant for the Higgsless case. Here,
we have to be concerned with the possibility that a light Higgs boson, as introduced in our 
present model, could alter significantly the decay dynamics of our $W^\pm_{1,2},Z_{1,2}$ states.
As we are not searching for a direct Higgs signal, we ought to only really look at  indirect 
Higgs effects on the total widths of our gauge bosons. Fig.~\ref{fig:gammas} shows the
typical corrections onto the Higgless width results due to the presence of a light composite Higgs boson with mass of 125 GeV, for our
usual $z$ choices. From such a figure,
one can see that  overall such Higgs induced effects 
are essentially negligible throughout, except for the case of the $Z_2$ state
at small $z$. Such modifications onset by the composite Higgs state will be accounted for
in the ensuing numerical analysis.

\subsection{Production phenomenology}

\begin{figure}[!t]
\begin{center}
\includegraphics[scale=0.8]{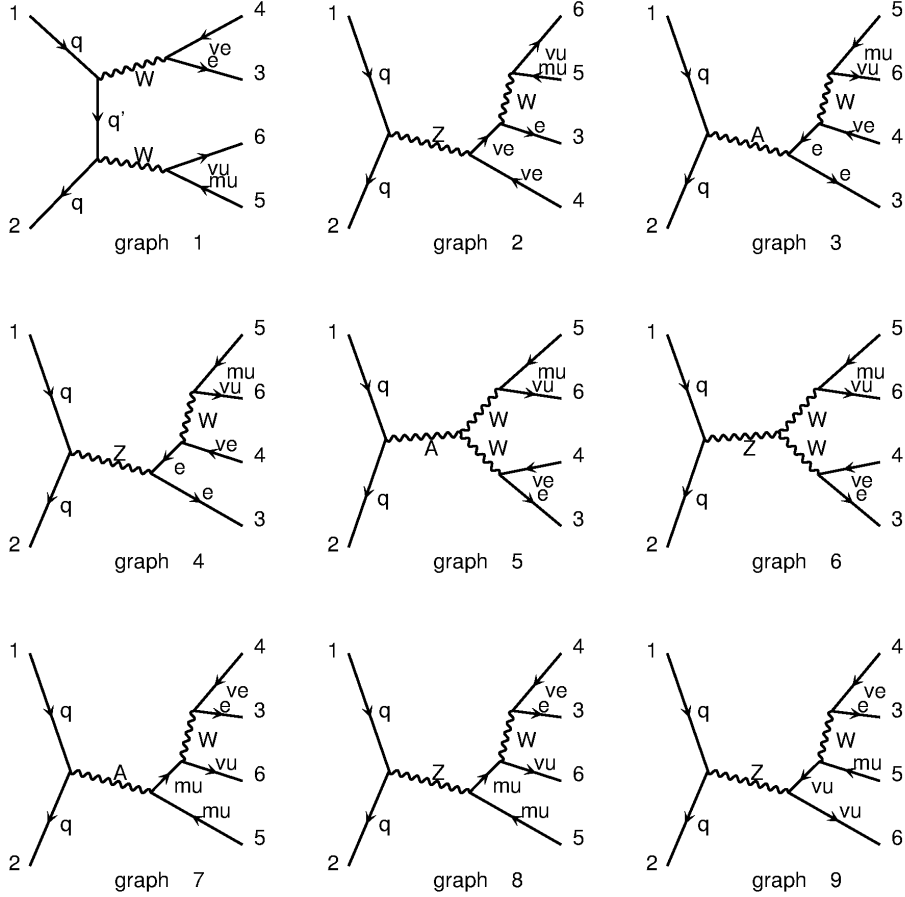}
\end{center}
\caption{Topologies of Feynman diagrams for the process in eq.~(\ref{eq:process}).
Here the labels $Z$ and $W$ refer to all possible gauge boson states of the model, neutral and charged,
respectively.} 
\protect{\label{fig:diagrams}}
\end{figure}

The codes exploited for our study of the LHC signatures
are based on helicity amplitudes, defined through either the PHACT module
\cite{Ballestrero:1999md} or the HELAS subroutines~\cite{Murayama:1992gi}, the latter assembled by means of MadGraph~\cite{Stelzer:1994ta}. 
The two independent subroutines were validated against each other. 
Further, the scattering amplitude for $gg\rightarrow WW\rightarrow e\mu+2\nu$, i.e., di-boson production via an $s$-channel scalar resonance,
was extracted from the codes used in \cite{Kunszt:1996yp}.
Two different phase space implementations were also adopted, an `ad-hoc one' (eventually used for event generation) and a 
`blind one' based on RAMBO \cite{Kleiss:1985gy}, again checked one against the other. 
VEGAS~\cite{Lepage:1977sw} was eventually used for the multi-dimensional numerical integrations.
The Matrix Elements (MEs) always account for all off-shellness effects
of the particles involved and were constructed starting from the topologies in Fig. 
\ref{fig:diagrams}, wherein the labels $Z$ and $W$ refer to any possible
combination of gauge bosons in our model. 
The Parton Distribution Functions (PDFs) used 
were CTEQ5L~\cite{Lai:1999wy}, with factorisation/renormalisation
scale set to $Q=\mu=\sqrt{\hat{s}}$. Initial state quarks have been taken
as massless, just like the final state leptons and neutrinos.

To calculate the cross section at the LHC for our model in the charged di-boson channel we consider three 
different set of cuts:  a `standard cuts'
scenario, a `soft cuts' scenario and a `hard cuts' scenario.\\
\vskip0.25cm
\textbf{Standard Cuts} ({\it St})~(some are inspired by Ref.~\cite{Moretti:1997ng}):
\begin{itemize}
\item $M_{ll}>180$~GeV (to avoid main SM contributions from the $Z$ and $WW$ peaks: notice that this cut is in fact hard-coded in our event generation) 
\item $|\eta_{l}|<3$ (this is a standard acceptance cut)
\item $p_T(l)>20$~GeV  (this is also a standard acceptance cut)
\item $E_{T}^{\rm miss}>50$~GeV (see Fig.~\ref{fig:distr_1000})
\item $\cos\phi_{ll}^T<-0.5$ (see Fig.~\ref{fig:cosll2})
\item $\cos\theta_{ll}<0$ (see Fig.~\ref{fig:cosll2})
\end{itemize}
where $M_{ll}^2=(p_e+p_\mu)^2$ is the invariant mass of the couple of charged leptons, $\eta_{l}$ is the pseudo-rapidity of the charged
leptons, $p_T(l)$ is the transverse momentum of the charged leptons, $E_{T}^{\rm miss}$~is the missing 
transverse energy, defined as
$(E_{T}^{\rm miss})^2=(p_T(e)+p_T(\mu))^2=(p_T(\nu_e)+p_T(\nu_\mu))^2$. Further, 
$\cos\phi_{ll}^T<-0.5$ is the cosine between the two leptons in the transverse plane whereas
$\cos\theta_{ll}$ is the cosine between the two leptons.
These are standard cuts, 
useful for a general purpose search.\\
\vskip0.25cm
\textbf{Soft Cuts} ({\it So}):
\begin{itemize}
\item $M_{ll}>180$~GeV 
\item $|\eta_{l}|<2$ (to exclude the regions where the difference with the SM is small)
\item $p_T(l)>20$~GeV
\item $E_{T}^{\rm miss}>50$~GeV
\item $P_T^{\rm max}(l)>180$~GeV  (see Fig.~\ref{fig:distr_1000})
\item $\cos\phi_{ll}^T<-0.5$
\item $\cos\theta_{ll}<0$ 
\end{itemize}
where $P_T^{\rm max}(l)={\rm max}(P_T(e),P_T(\mu))$. These cuts are studied to further suppress the SM, 
leaving however not too small a signal cross section.\\
\vskip0.25cm
\textbf{Hard Cuts} ({\it Ha}):
\begin{itemize}
\item $M_{ll}>220$~GeV 
\item $|\eta_{l}|<1.5$ 
\item $p_T(l)>20$~GeV 
\item $E_{T}^{\rm miss}>220$~GeV 
\item $P_T^{\rm max}>220$~GeV 
\item $\cos\phi_{ll}^T<-0.5$ 
\item $\cos\theta_{ll}<0$ 
\end{itemize}
which represent a general tightening of (some of) the previous ones, at a further cost to the signal.
We will be using one or more of such cut combinations to explore the parameter space of our model, via
 di-boson production, after the preliminary exercise of displaying
typical cross sections (both inclusive and exclusive) for it. For the case of DY processes, we instead refer 
the reader to Refs.~\cite{Accomando:2011eu,Accomando:2011xi,Accomando:2010ir,Accomando:2008jh,Accomando:2008dm}.

\subsection{Distributions}
\begin{figure}[!t]
\begin{center}
\vspace{-.8cm}
\unitlength1.0cm
\begin{picture}(7,10)
\put(-4.3,2.7){\epsfig{file=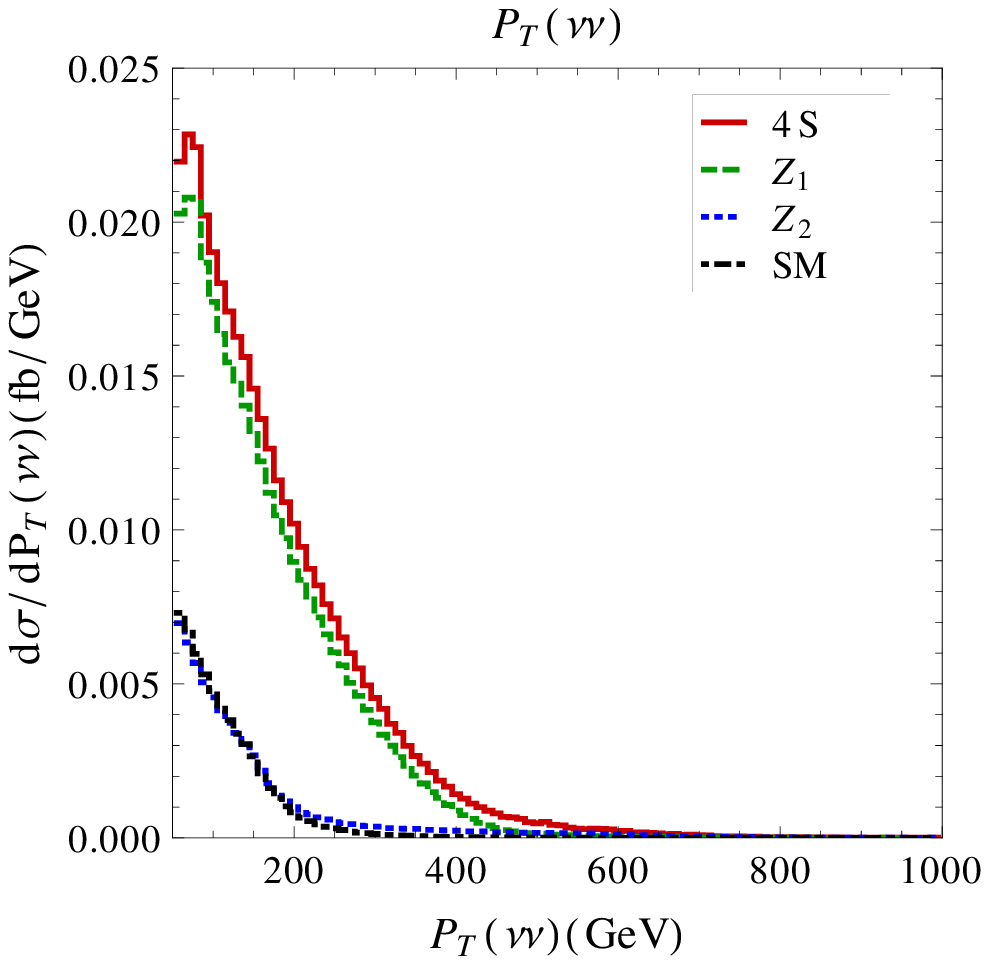,width=7.5cm}}
\put(3.5,2.7){\epsfig{file=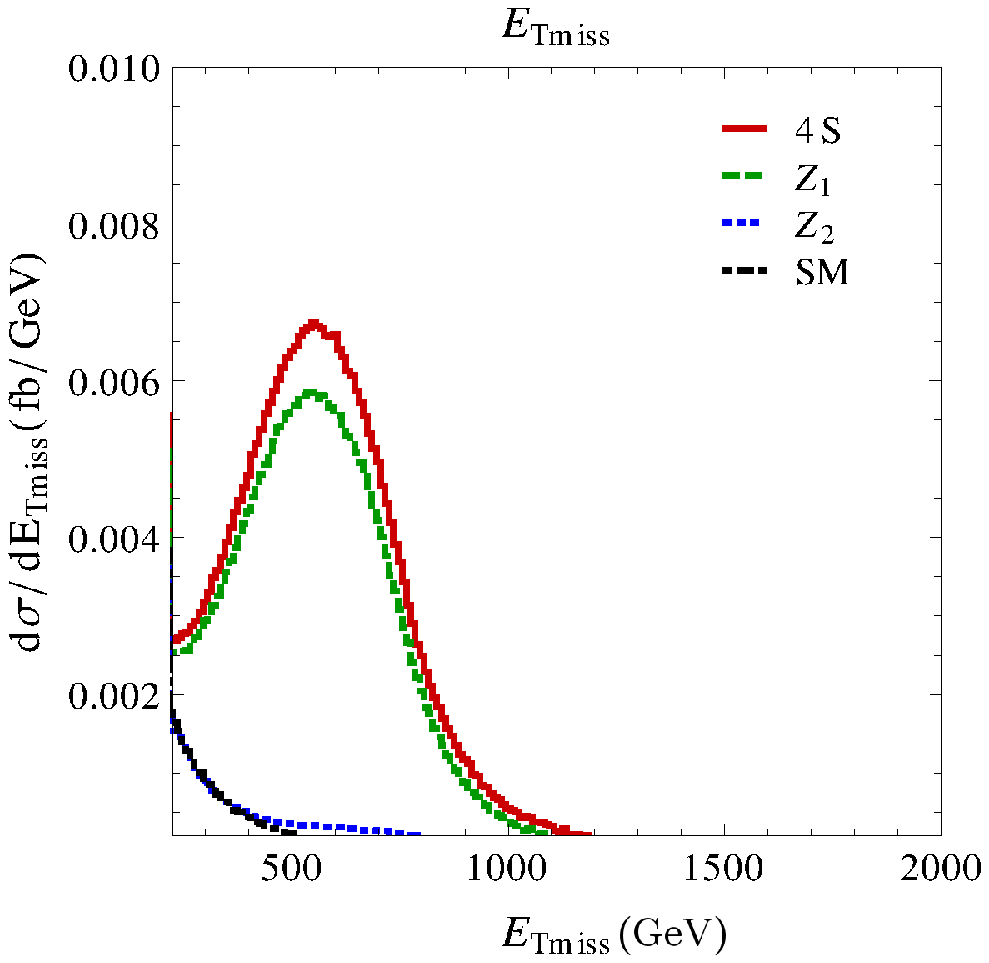,width=7.5cm}}
\put(-4.3,-5){\epsfig{file=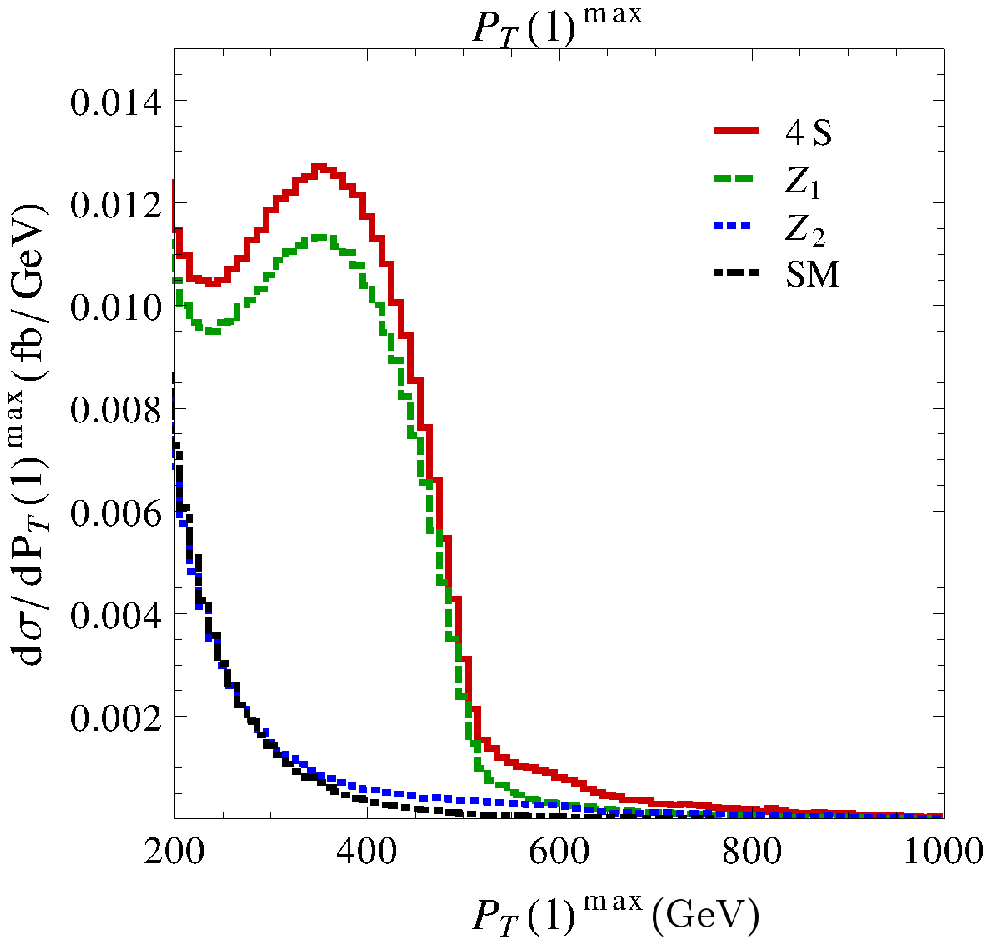,width=7.5cm}}
\put(3.5,-5){\epsfig{file=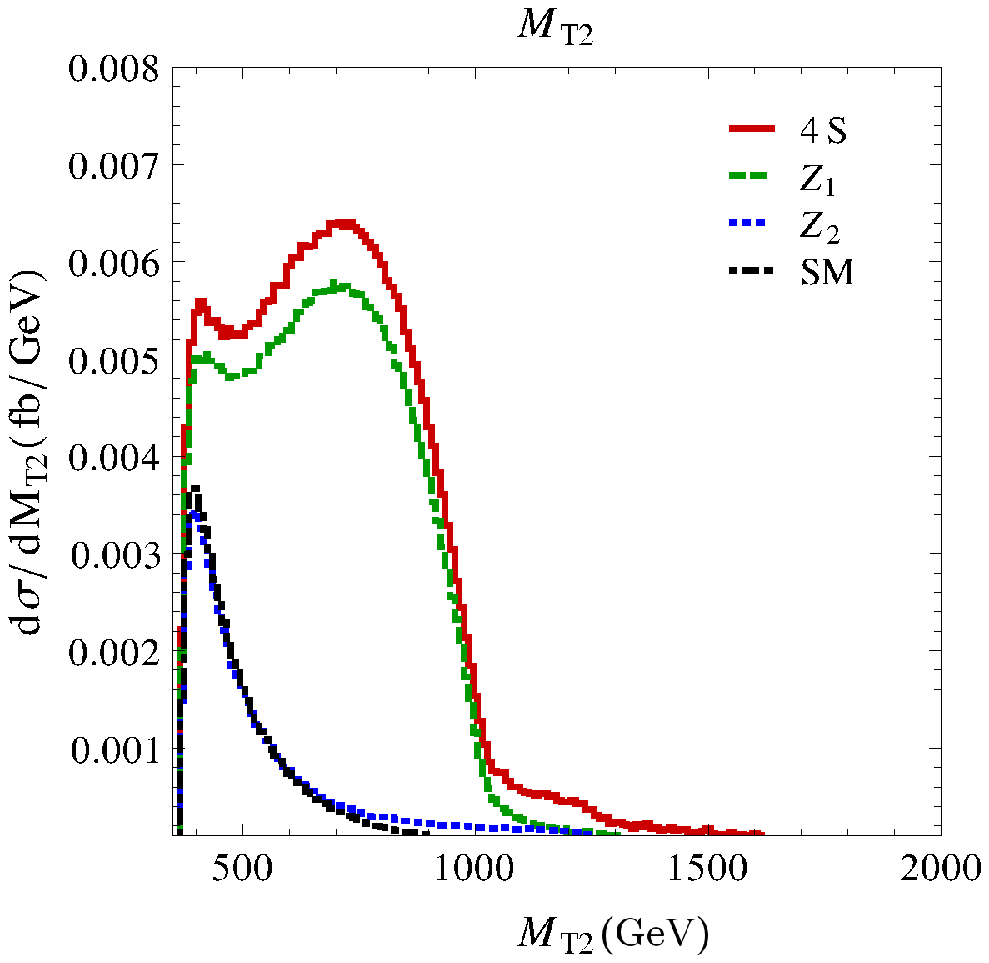,width=7.5cm}}
\end{picture}
\end{center}
\vskip 4.cm \caption{Differential cross sections pertaining to the di-boson process for the 4-site model, for 
$M_1$=1~TeV and $z=0.8$. We choose the maximal not excluded value for
$a_{W_1}$. Here, $P_T(\nu\nu)$ is the transverse momentum of two neutrinos in the transverse plane as defined in the text
whereas $M_{T2}$ is the transverse mass as defined in \protect{Ref.~\cite{AA:2012ks}}.  
The red-solid curve represents the full 4-site model, 
the green-dashed(blue-dotted) curve represents the $Z_1$($Z_2$) contribution alone and the black-dotted-dashed curve is the SM. {\it So} cuts were applied.
}
\label{fig:distr_1000}
\end{figure}

\begin{figure}[!t]
\begin{center}
\unitlength1.0cm
\begin{picture}(7,4)
\put(-4.5,-4){\epsfig{file=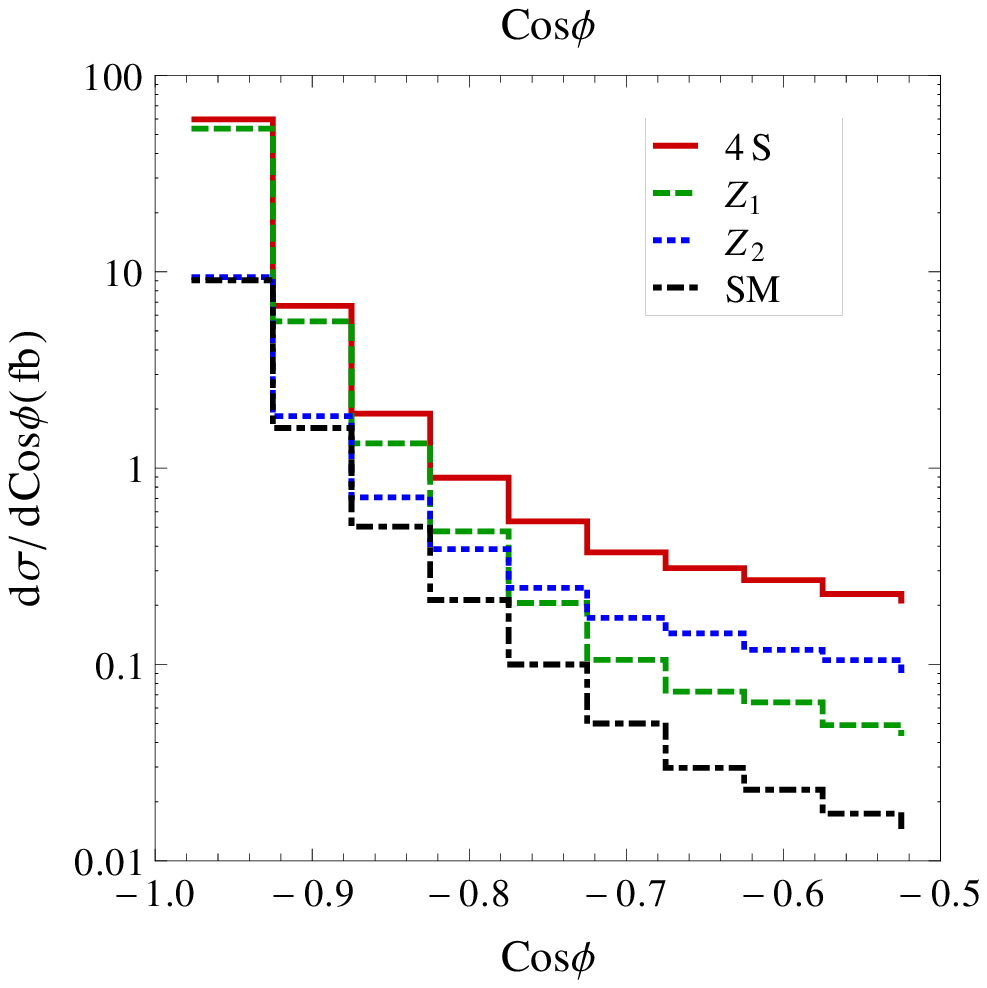,width=7.5cm}}
\put(3.3,-4){\epsfig{file=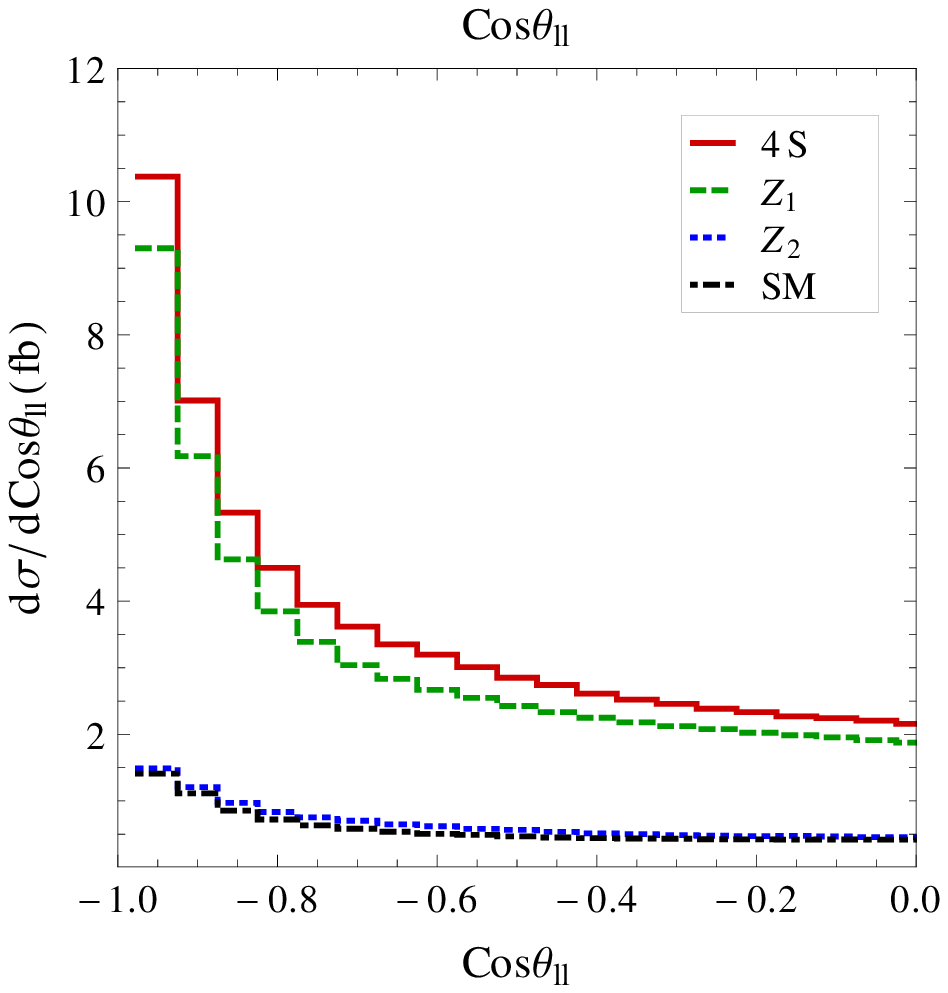,width=7.5cm}}
\end{picture}
\end{center}
\vskip 3.cm
\caption{Differential cross sections pertaining to the di-boson process for the 4-site model, for 
$M_1$=1~TeV and $z=0.8$. We choose the maximal  not excluded value for
$a_{W_1}$. {\it So} cuts are applied.
The red-solid curve represents the full 4-site model, 
the green-dashed(blue-dotted) curve represents the $Z_1$($Z_2$) contribution alone and the black-dotted-dashed curve is the SM. {\it So} cuts were applied.
} 
\label{fig:cosll2}
\end{figure}

\begin{figure}[!t]
\begin{center}
\vspace{-.8cm}
\unitlength1.0cm
\begin{picture}(7,10)
\put(-4.3,2.7){\epsfig{file=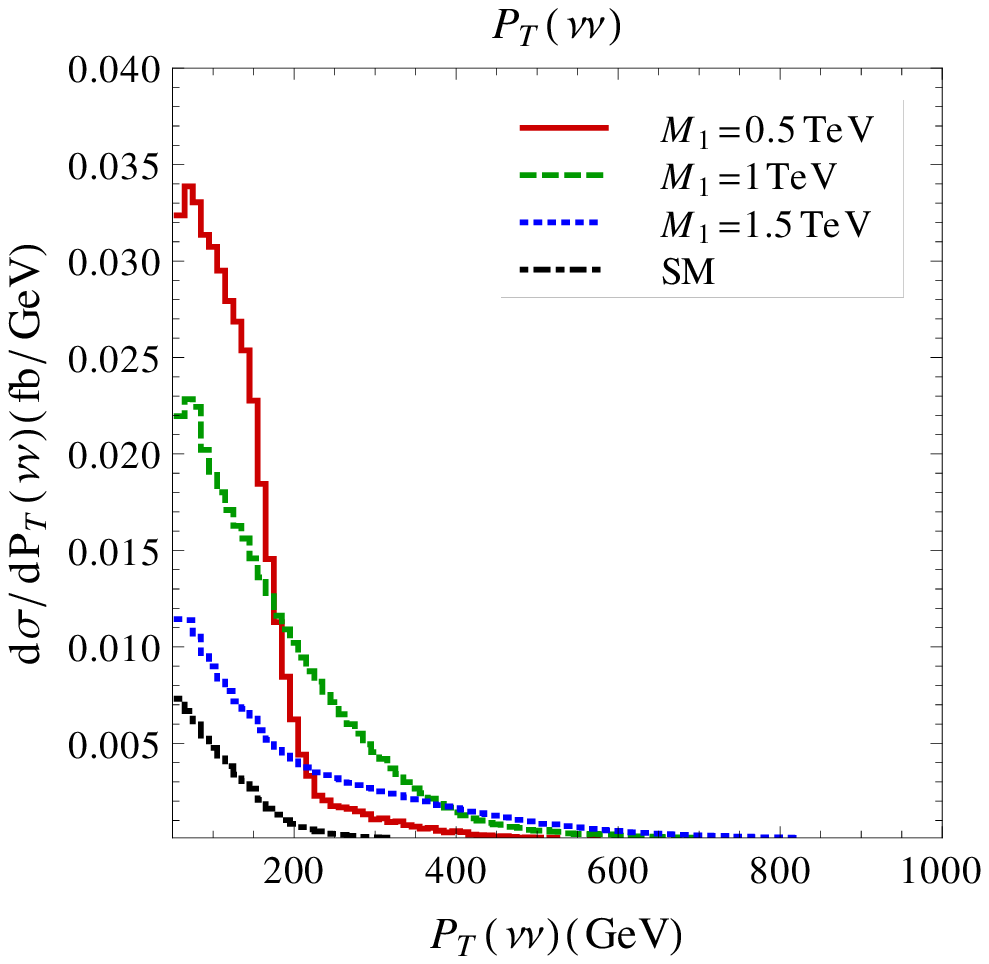,width=7.1cm}}
\put(3.5,2.7){\epsfig{file=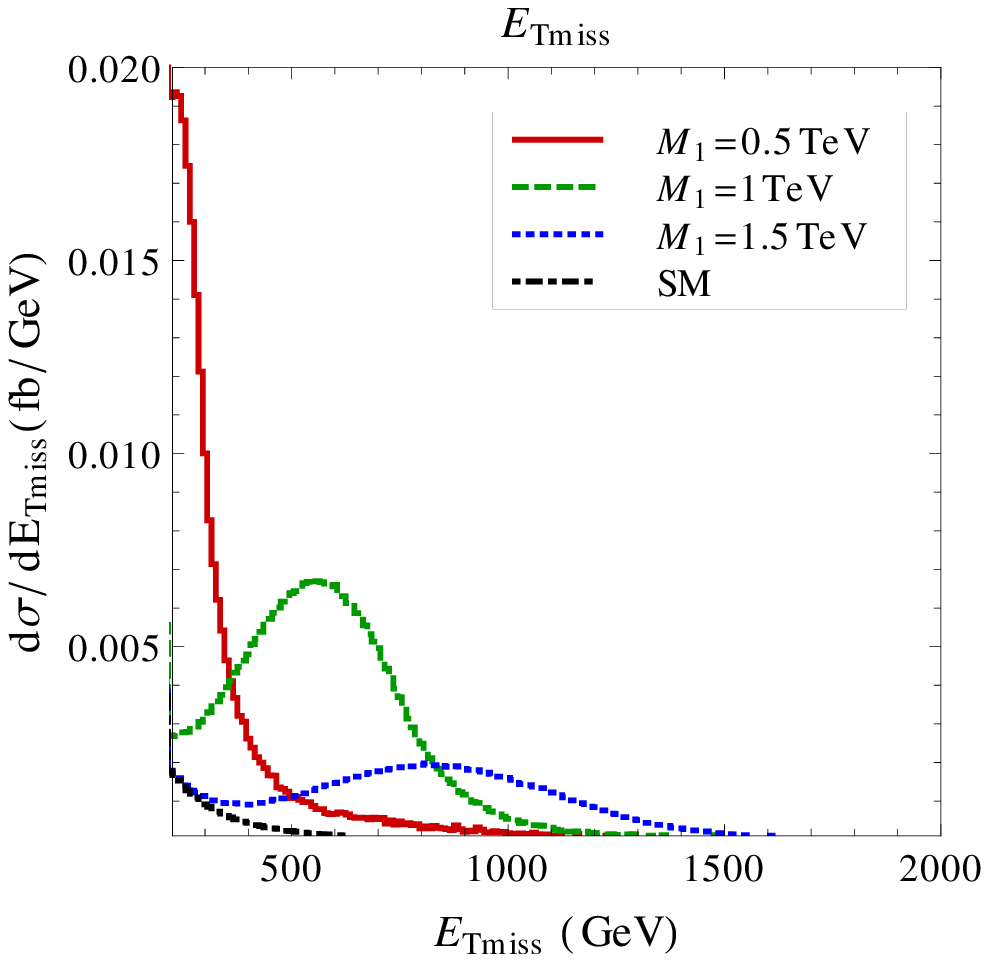,width=7.5cm}}
\put(-4.3,-5){\epsfig{file=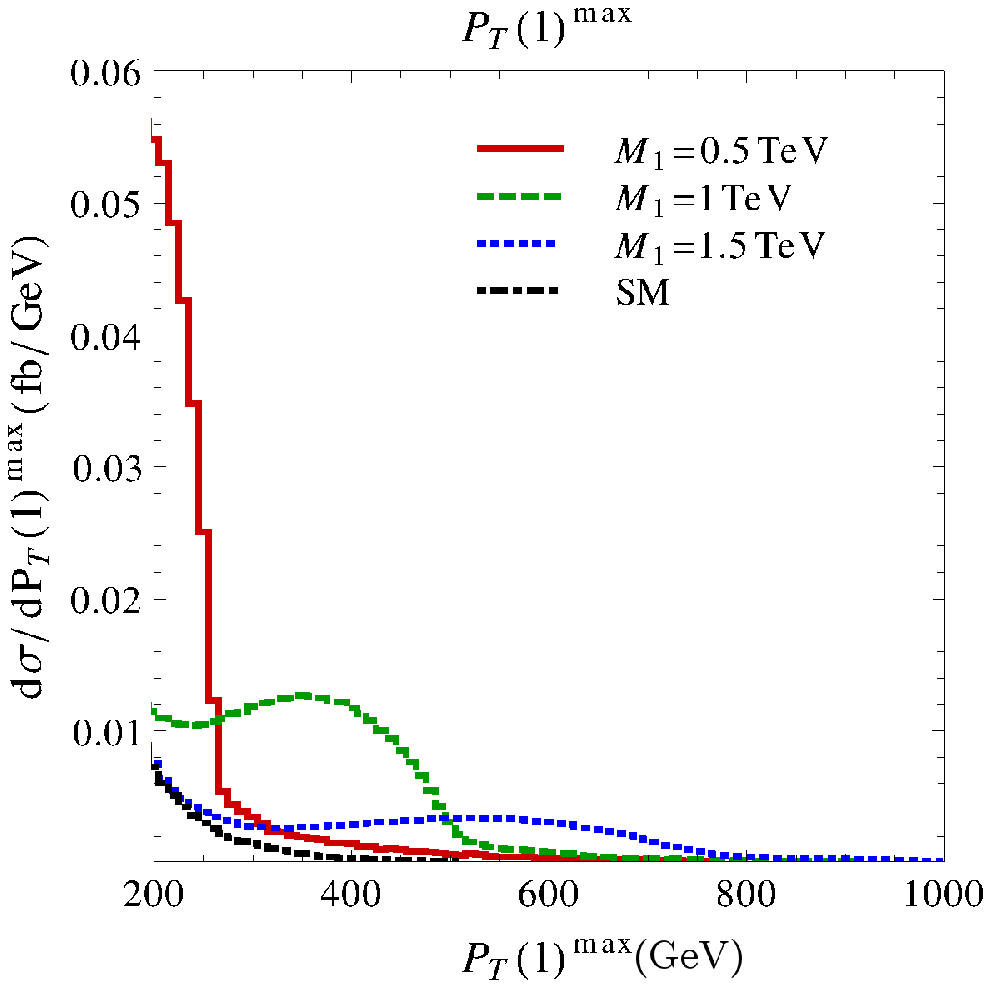,width=7.5cm}}
\put(3.5,-5){\epsfig{file=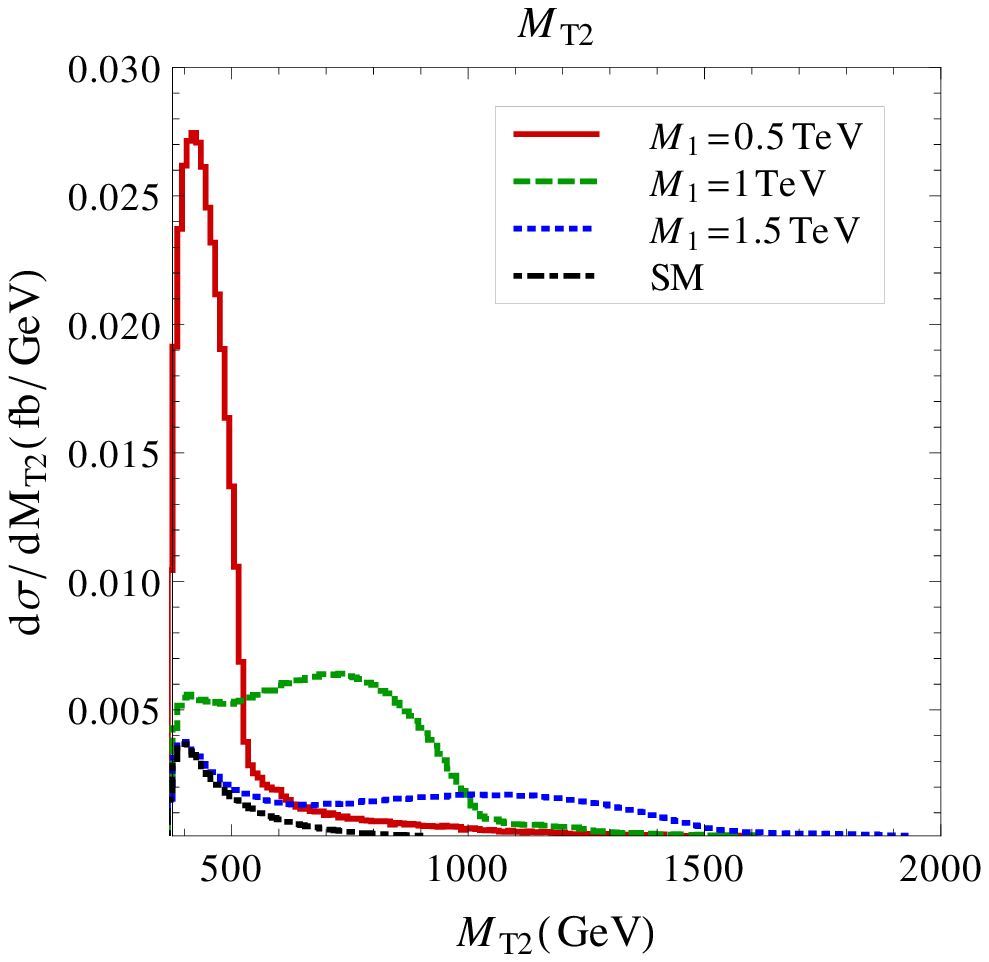,width=7.5cm}}
\end{picture}
\end{center}
\vskip 4.cm \caption{Differential cross sections pertaining to the di-boson process for the 4-site model, for 
$M_1$=0.5, 1 and 1.5~TeV and $z=0.8$. We choose the maximal  not excluded  value for
$a_{W_1}$.
Some relevant distributions concerning the di-boson process for the 4-site model, for $M_1$=0.5 (red-solid), 1 (green-dashed), 1.5 (blue-dotted)~TeV. The black-dot-dashed curve is the SM. All observables have been previously defined.  {\it So} cuts are applied.  
}
\label{fig:distr_z08}
\end{figure}

Before exploring the full parameter space it is useful to consider some total rates and differential distributions for
process ({\ref{eq:process}}), in a such way to understand the 
relevant new contributions to the cross section. (Incidentally, we ought to notice at this point that process (\ref{eq:Higgs}), despite
giving fully inclusive production cross sections of order tens of fb at all energy stages of the CERN machine, after any of the above sets of cuts is applied, turns out to fall under observability limits for
all considered luminosities, so that we neglect considering it further 
in our analysis.) In Fig.~\ref{fig:distr_1000} we consider 
$M_1=1$~TeV and the maximal allowed value for $a_{W_1}$ for $z=0.8$  (see Fig. \ref{fig:EWPT_a1-a2}),
and we display four relevant observables (we use the \textit{So} cuts here). Herein, in order to better understand the role of the single neutral resonances 
in $s$-channel, we show also the contribution from 
the $Z_1$ and $Z_2$ resonances separately. It is quite clear that the second resonance ($Z_2$) is almost invisible and does not contribute 
to the total cross section. This is due to the fact that the trilinear gauge vertex ($Z_2WW$) is strongly suppressed due to the axial characteristics
of the $Z_2$ state, so the only visible resonance is the lighter one. Remarkably, this is a completely different scenario from the DY one, in which the heavy state contribution to DY in both the Charged Current
(CC) and Neutral Current (NC) case is the dominant one and often (especially in the former case) covers also the signal of the lighter one. This fact renders the di-boson
channel a very valuable process to exploit in order to complement the scope of the DY one, so that both modes can be taken together to effectively 
cooperate in allowing one to see the typical multi-resonance structure of the gauge sector of the 4-site model.
In addition to observables already introduced when defining the cuts, we also use the following additional ones:
\be
P_T(\nu\nu)=\sqrt{p_T(e)^2+p_T(\mu)^2},\quad M_{T2}=p_T(e)+p_T(\mu)+E_T^{ll},
\ee
where
\be
E_T^{ll}=p(e)^0+p(\mu)^0,
\ee
which were not adopted for the final selections, yet they show some sizable differences between
signal and background. 
In Fig.~\ref{fig:cosll2} we present the angular distributions used for our selection cuts, for the
purpose of motivating the latter (the behaviour of the curves established towards the right
end of the angular intervals plotted is maintained beyond it too). Finally, in
Fig.~\ref{fig:distr_z08} we show the same relevant observables of
Fig.~\ref{fig:distr_1000}, considering three different mass scenarios (still for $z=0.8$), proving
that they are generally effective independently of the mass values of the resonances.

\subsection{Exploring the parameter space}
In this section we present some numerical values for the cross section at the LHC with, initially, 7 TeV, for some benchmark points in the parameter space, defined by 
 five sets of masses for each one of the four chosen $z$ values. For each mass we consider the maximal allowed value for $a_{W_1}$. The mass values are
$M_1=$0.5, 0.75, 1, 1.5 and 2~TeV for $z$ from 0.4 to 0.8. For $z=0.95$ the lower value for the mass is excluded, so we consider instead
the scenario $M_1=1.7$~TeV. In Tabs.~\ref{tab:z04}, \ref{tab:z06}, \ref{tab:z08} and  \ref{tab:z95} we show the cross sections for such
scenarios, for all our cut choices and including also a `{No Cuts}' reference scenario (as defined in the captions).
In Tabs.~\ref{tab:z04_Stat}, \ref{tab:z06_Stat}, \ref{tab:z08_Stat} and  \ref{tab:z95_Stat} we present instead the statistical significance $S$, defined
as in eq.~(\ref{eq:signi}), considering a luminosity of 10~fb$^{-1}$. Here, we note that, for almost all masses and $z$ values considered, the \textit{So}
cuts allow for the largest statistical significances.
Therefore, we decided to use the \textit{So} cuts to explore the full parameter space of the 4-site model with
larger data samples, in particular, we consider both
the actual (8 TeV with 5~fb$^{-1}$) and future (8 and 14~TeV with 15~fb$^{-1}$) LHC scenarios. In Tabs.~\ref{tab:LHC8}-\ref{tab:LHC14_Stat}
are listed the ensuing cross sections and statistical significances. 

\begin{table}[!htb]
\begin{center}
\begin{tabular}{||c|c|c|c|c||}
\hline \hline
$z=0.4$, $M_1$ [TeV]&No Cuts$^1$ (10.5) [fb]& {\it St} (1.7) [fb]& {\it So} (0.58) [fb]& {\it Ha} (0.025) [fb]\\
\hline 
\hline
 0.5&21.2 & 6.09 & 3.16 & 0.17 \\
\hline
0.75& 16.7 & 5.14 & 3.66 & 0.39 \\
\hline
1& 13.7 & 3.84 & 2.61 & 0.52 \\
\hline
1.5& 11.2 & 2.55 & 1.78 & 0.30 \\
\hline
2& 10.7 & 1.83 & 0.74 & 0.12\\
\hline\hline
\end{tabular}
\end{center}
\caption{Cross section for the 4-site model considering the process of eq.~(\ref{eq:process}) at the 7~TeV LHC for five different mass values and $z=0.4$.
 The values in parenthesis refer to the SM.  
$^1$The only cut is on the invariant mass of the charged leptons and is set to 180 GeV.}
\label{tab:z04}
\end{table}
\begin{table}[!htb]
\begin{center}
\begin{tabular}{||c|c|c|c|c||}
\hline \hline
$z=0.4$, $M_1$ [TeV]&No Cuts$^1$ & {\it St} & {\it So} & {\it Ha} \\
\hline 
\hline
 0.5& 10.4 & 10.6 & 11.4 & 1.5 \\
\hline
 0.75& 6.0 & 8.3 & 13.6 & 3.7 \\
\hline
 1& 3.1 & 5.2 & 9.0 & 5.0 \\
\hline
 1.5& 0.7 & 2.1 & 5.4 & 2.8 \\
\hline
  2&0.2 & 0.3 & 0.9 & 1.0\\
\hline\hline
\end{tabular}
\end{center}
\caption{Statistical significance ($S$) for the 4-site model considering the process of eq.~(\ref{eq:process}) at the 7 TeV 
LHC for different mass values and $z=0.4$, for L=10~fb$^{-1}$.
$^1$The only cut is on the invariant mass of the charged leptons and is set to 180 GeV.}
\label{tab:z04_Stat}
\end{table}
\begin{table}[!htb]
\begin{center}
\begin{tabular}{||c|c|c|c|c||}
\hline \hline
$z=0.6$, $M_1$ [TeV]&No Cuts$^1$ (10.5) [fb]& {\it St} (1.7) [fb]& {\it So} (0.58) [fb]& {\it Ha} (0.025) [fb]\\
\hline \hline
0.5& 25.7 & 7.82 & 4.25 & 0.22 \\
\hline
0.75& 19.5 & 6.67 & 4.97 & 0.61 \\
\hline
1& 14.8 & 4.56 & 3.39 & 0.75 \\
\hline
1.5& 11.4 & 2.52 & 1.40 & 0.41 \\
\hline
2& 10.6 & 1.87 & 0.79 & 0.14\\
\hline\hline
\end{tabular}
\end{center}
\caption{Same as Tab. \ref{tab:z04} for $z=0.6$.}
\label{tab:z06}
\end{table}
\begin{table}[!htb]
\begin{center}
\begin{tabular}{||c|c|c|c|c||}
\hline \hline
$z=0.6$, $M_1$ [TeV]&No Cuts$^1$ & {\it St} & {\it So} & {\it Ha} \\
\hline 
\hline
0.5& 14.8 & 14.8 & 16.2 & 2.0 \\
\hline
0.75& 8.8 & 12.0 & 19.3 & 5.9 \\
\hline
1& 4.2 & 6.9 & 12.4 & 7.3 \\
\hline
1.5& 0.9 & 2.0 & 3.8 & 3.9 \\
\hline
2& 0.2 & 0.4 & 1.1 & 1.2\\
\hline\hline
\end{tabular}
\end{center}
\caption{Same as Tab. \ref{tab:z04_Stat} for $z=0.6$.}
\label{tab:z06_Stat}
\end{table}
\begin{table}[!htb]
\begin{center}
\begin{tabular}{||c|c|c|c|c||}
\hline \hline
$z=0.8$, $M_1$ [TeV]&No Cuts$^1$ (10.5) [fb]& {\it St} (1.7) [fb]& {\it So} (0.58) [fb]& {\it Ha} (0.025) [fb]\\
\hline 
0.5&23.7 & 6.85  & 3.78 &0.22 \\
\hline 
1&15.2 & 4.69  & 3.55 &0.87 \\
\hline
1.5& 12.1& 2.97 & 1.91 & 0.68\\
\hline
1.7& 11.4& 2.51  & 1.38 &0.48 \\
\hline
2& 10.7& 2.03  & 0.91 &0.23 \\
\hline\hline
\end{tabular}
\end{center}
\caption{Same as Tab. \ref{tab:z04} for $z=0.8$.}
\label{tab:z08}
\end{table}
\begin{table}[!htb]
\begin{center}
\begin{tabular}{||c|c|c|c|c||}
\hline \hline
$z=0.8$, $M_1$ [TeV]&No Cuts$^1$ & {\it St} & {\it So} & {\it Ha} \\
\hline 
\hline
0.5& 12.9 & 12.5 & 14.1 & 2.0 \\
\hline
1& 4.6 & 7.2 & 13.1 & 8.5 \\
\hline
1.5& 1.6 & 3.1 & 6.0 & 6.6 \\
\hline
1.7& 0.9 & 2.0 & 3.7 & 4.6\\
\hline
2& 0.2 & 0.8 & 1.6 & 2.1 \\
\hline\hline
\end{tabular}
\end{center}
\caption{Same as Tab.  \ref{tab:z04_Stat} for $z=0.8$.}
\label{tab:z08_Stat}
\end{table}
\begin{table}[!htb]
\begin{center}
\begin{tabular}{||c|c|c|c|c||}
\hline \hline
$z=0.95$, $M_1$ [TeV]&No Cuts$^1$ (10.5) [fb]& {\it St} (1.7) [fb]& {\it So} (0.58) [fb]& {\it Ha} (0.025) [fb]\\
\hline \hline
0.75& 14.8 & 3.97 & 2.67 & 0.44 \\
\hline
1& 12.0& 2.62 & 1.54 & 0.41 \\
\hline
1.5& 11.1 & 2.19 & 1.09 & 0.34 \\
\hline
1.7& 11.4 & 2.28 & 1.17 & 0.45 \\
\hline
2& 11.1& 2.14 & 1.07 & 0.41\\
\hline\hline
\end{tabular}
\end{center}
\caption{Same as Tab. \ref{tab:z04} for $z=0.95$.}
\label{tab:z95}
\end{table}
\begin{table}[!htb]
\begin{center}
\begin{tabular}{||c|c|c|c|c||}
\hline \hline
$z=0.95$, $M_1$ [TeV]&No Cuts$^1$ & {\it St} & {\it So} &{\it Ha} \\
\hline 
\hline
0.75& 4.2 & 5.5 & 9.3& 4.2 \\
\hline
1& 1.5 & 2.2 & 4.4 & 3.9 \\
\hline
1.5& 0.6 & 1.2 & 2.4 & 3.2 \\
\hline
1.7& 0.9 & 1.4 & 2.8 & 4.3 \\
\hline
2& {1.0} & 1.1 & 2.3 & 3.9\\
\hline\hline
\end{tabular}
\end{center}
\caption{Same as Tab. \ref{tab:z04_Stat} for $z=0.95$.}
\label{tab:z95_Stat}
\end{table}
\begin{table}[!htb]
\begin{center}
\begin{tabular}{||c|c|c|c|c|c|c||}
\hline \hline
 $M_1$ [TeV]&0.5&0.75&1&1.5&1.7&2\\
\hline \hline
$z=0.4$&  4.15 & 4.08 & 3.58 & 1.7 & 1.31 & 1.00 \\
\hline
$z=0.6$&  5.53 & 6.89 & 5.24 & 2.36 & 1.62 & 1.15 \\
\hline
$z=0.8$&  4.73 & 5.32 & 4.72 & 2.9 & 2.21 & 1.43 \\
\hline
$z=0.95$& -& 3.56 & 2.06 & 1.6 & 1.83 & 1.74\\
\hline\hline
\end{tabular}
\end{center}
\caption{Cross sections (in [fb]) for the 4-site model considering the process of eq.~(\ref{eq:process}) at the 8~TeV LHC for different mass and $z$ values.
 The value for the SM is 0.73~fb.  We consider {\it So} cuts only.}
\label{tab:LHC8}
\end{table}
\begin{table}[!htb]
\begin{center}
\begin{tabular}{||c|c|c|c|c|c|c||}
\hline \hline
$M_1$ [TeV]&0.5&0.75&1&1.5&1.7&2\\
\hline 
\hline
$z=0.4$&  15.5 & 15.2 & 12.9 & 4.4 & 2.6 & 1.2 \\
\hline
$z=0.6$& 21.8 & 27.9 & 20.4 & 7.4 & 3.9 & 1.9 \\
\hline
$z=0.8$&  18.1 & 20.8 & 18.1 & 9.8 & 6.7 & 3.2 \\
\hline
$z=0.95$&  12.8 & 12.8 & 6.0 & 3.9 & 5.0 & 4.6\\
\hline\hline
\end{tabular}
\end{center}
\caption{Statistical significance ($S$) for the 4-site model considering the process of eq.~(\ref{eq:process}) at the 8~TeV LHC for different mass and $z$ values,
 for L=15~fb$^{-1}$.  We consider {\it So} cuts only.}
\label{tab:LHC8_Stat}
\end{table}
\begin{table}[!htb]
\begin{center}
\begin{tabular}{||c|c|c|c|c|c|c||}
\hline \hline
 $M_1$ [TeV]&0.5&0.75&1&1.5&1.7&2\\
\hline \hline
$z=0.4$& 10.4 & 12.5 & 10.4 & 6.26 & 5.09 & 3.86 \\
\hline
$z=0.6$& 12.7 & 17.7 & 15.3 & 9.87 & 6.82 & 5.15 \\
\hline
$z=0.8$& 11.1 & 13.3 & 14.9 & 13.3 & 11.4 & 8.38 \\
\hline
$z=0.95$& - & 9.36 & 8.8 & 6.95 & 5.1 & 4.56 \\
\hline\hline
\end{tabular}
\end{center}
\caption{Cross sections (in [fb]) for the 4-site model considering the process of eq.~(\ref{eq:process}) at the 14~TeV LHC for different mass and $z$ values.
 The value for the SM is 1.55~fb.  We consider the {\it So} cuts only.}
\label{tab:LHC14}
\end{table}
\begin{table}[!htb]
\begin{center}
\begin{tabular}{||c|c|c|c|c|c|c||}
\hline \hline
$M_1$ [TeV]&0.5&0.75&1&1.5&1.7&2\\
\hline 
\hline
$z=0.4$&  22.5 & 27.8 & 22.5 & 12.0 & 9.0 & 5.9 \\
\hline
$z=0.6$&  28.3 & 41.0 & 34.9 & 21.1 & 13.4 & 9.1 \\
\hline
$z=0.8$&  24.3 & 29.8 & 33.9 & 29.8 & 25.0 & 17.3 \\
\hline
$z=0.95$&  - & 19.8 & 18.4 & 13.7 & 9.02 & 7.6\\
\hline\hline
\end{tabular}
\end{center}
\caption{Statistical significance ($S$) for the 4-site model considering the process of eq.~(\ref{eq:process}) at the 14~TeV LHC for different mass and $z$ values,
 for L=10~fb$^{-1}$.  We consider {\it So} cuts only.}
\label{tab:LHC14_Stat}
\end{table}

\subsection{Exclusion and discovery bounds}
\begin{figure}[!t]
\begin{center}
\vspace{-.8cm}
\unitlength1.0cm
\begin{picture}(7,10)
\put(-4.3,2.7){\epsfig{file=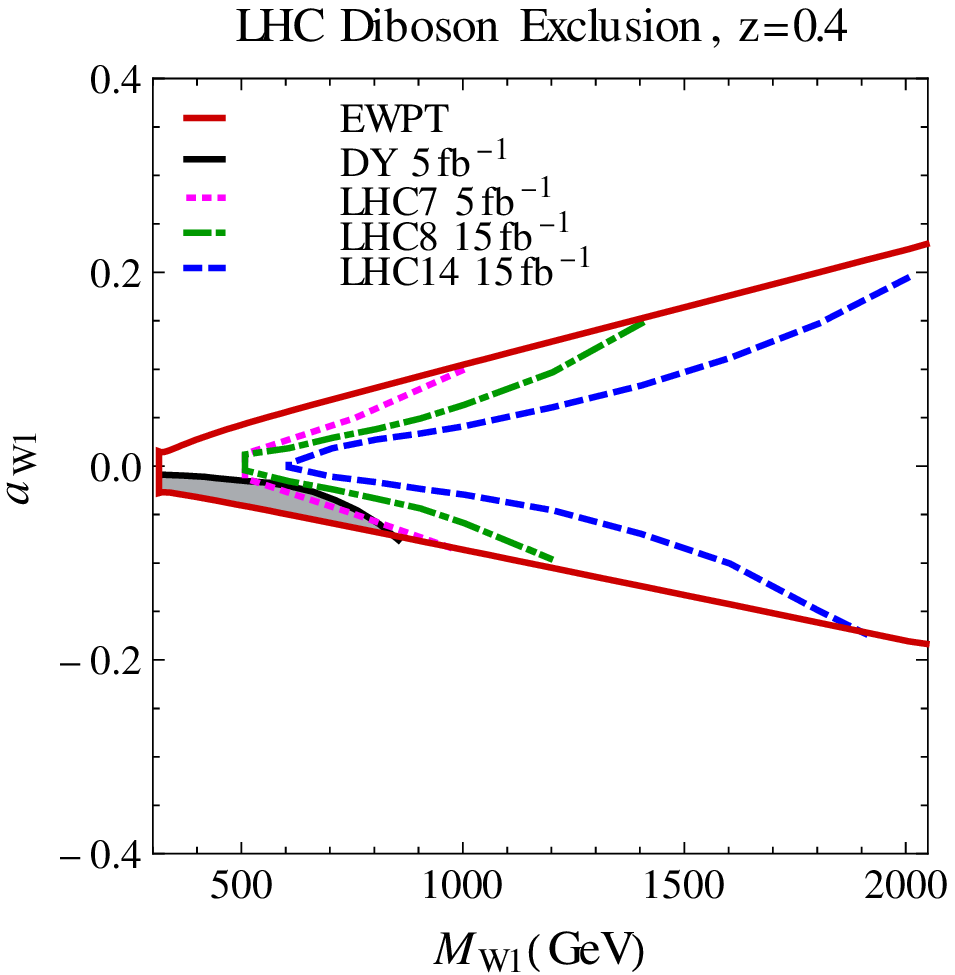,width=7.1cm}}
\put(3.5,2.7){\epsfig{file=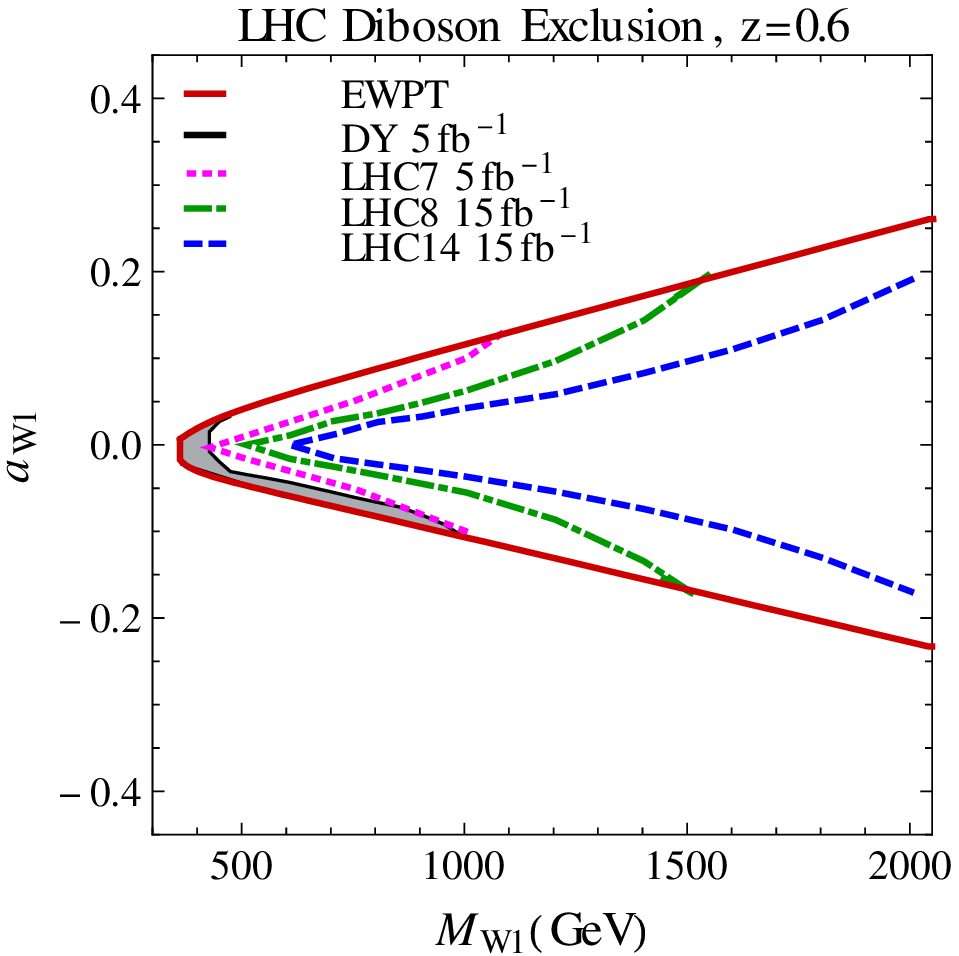,width=7.5cm}}
\put(-4.3,-5){\epsfig{file=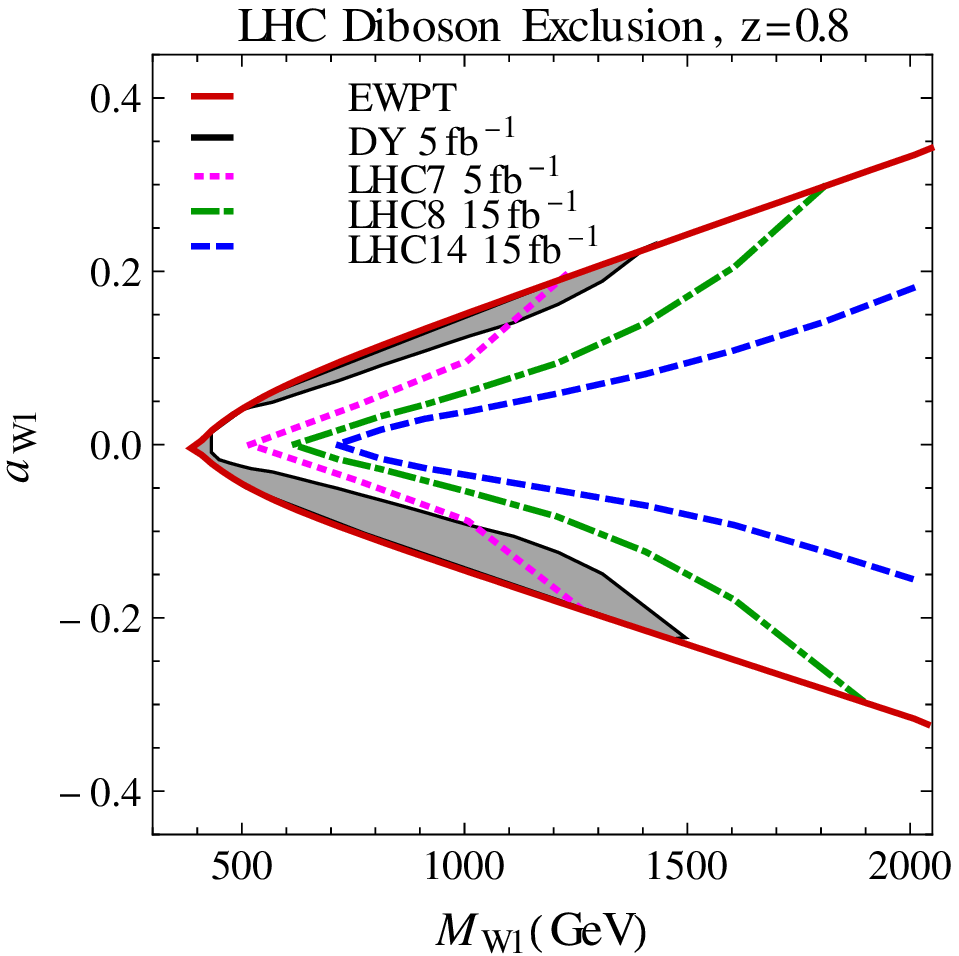,width=7.5cm}}
\put(3.5,-5){\epsfig{file=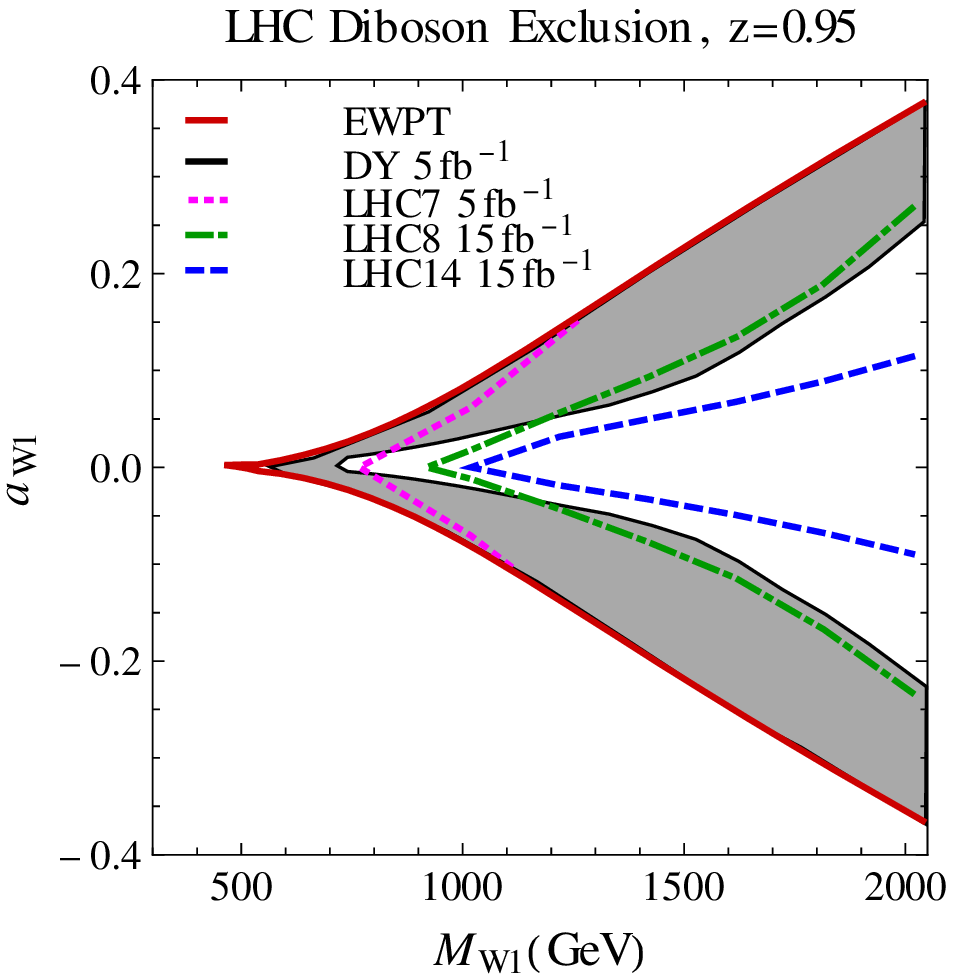,width=7.5cm}}
\end{picture}
\end{center}
\vskip 4.cm \caption{Shown are the 95\% CL exclusion limits from the LHC on the 4-site model: here, we consider the di-boson channel for the usual four values of the $z$ parameter. 
We assume all possible LHC setups considered.}
\label{fig:E_dib}
\end{figure}

\begin{figure}[!t]
\begin{center}
\vspace{-.8cm}
\unitlength1.0cm
\begin{picture}(7,10)
\put(-4.3,2.7){\epsfig{file=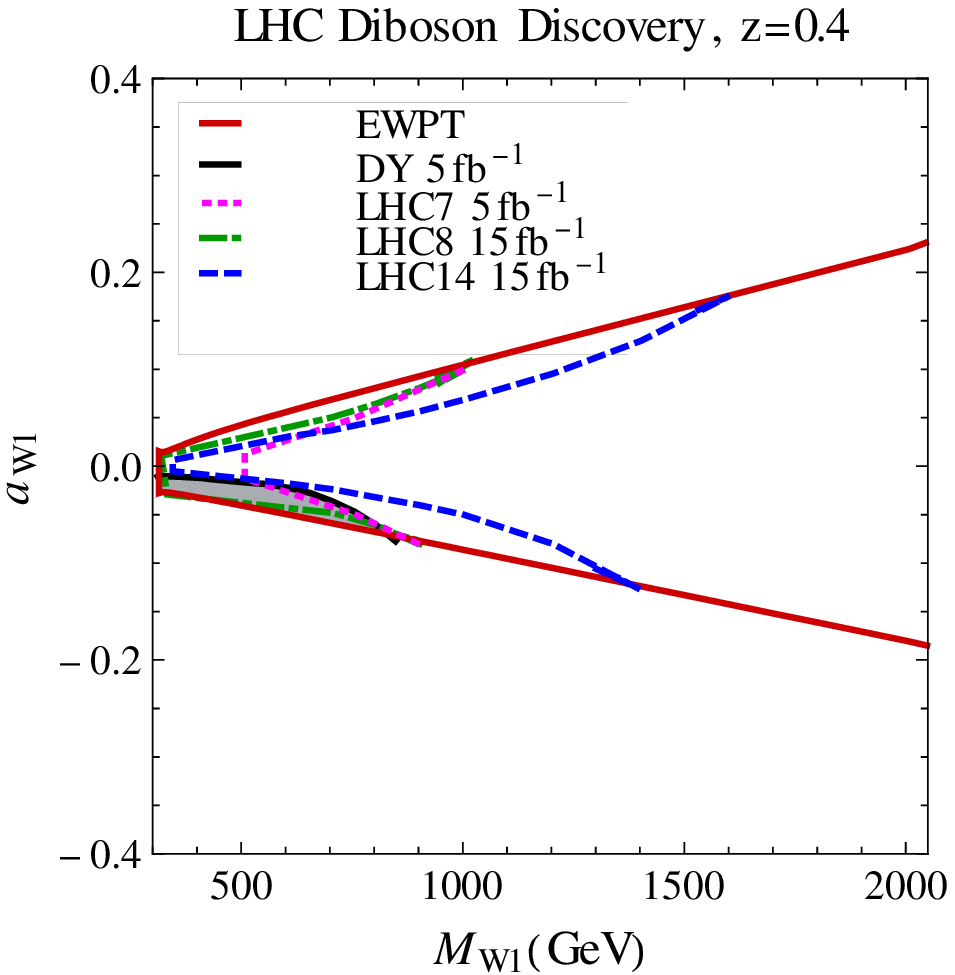,width=7.1cm}}
\put(3.5,2.7){\epsfig{file=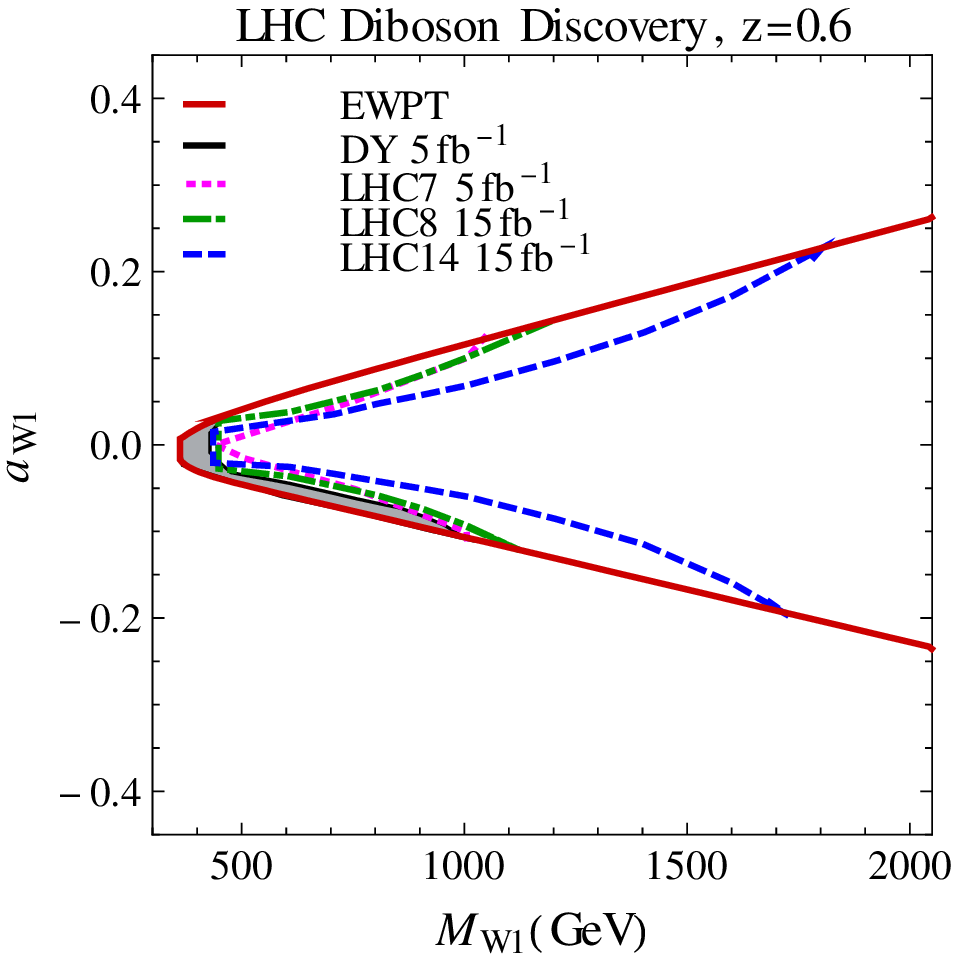,width=7.5cm}}
\put(-4.3,-5){\epsfig{file=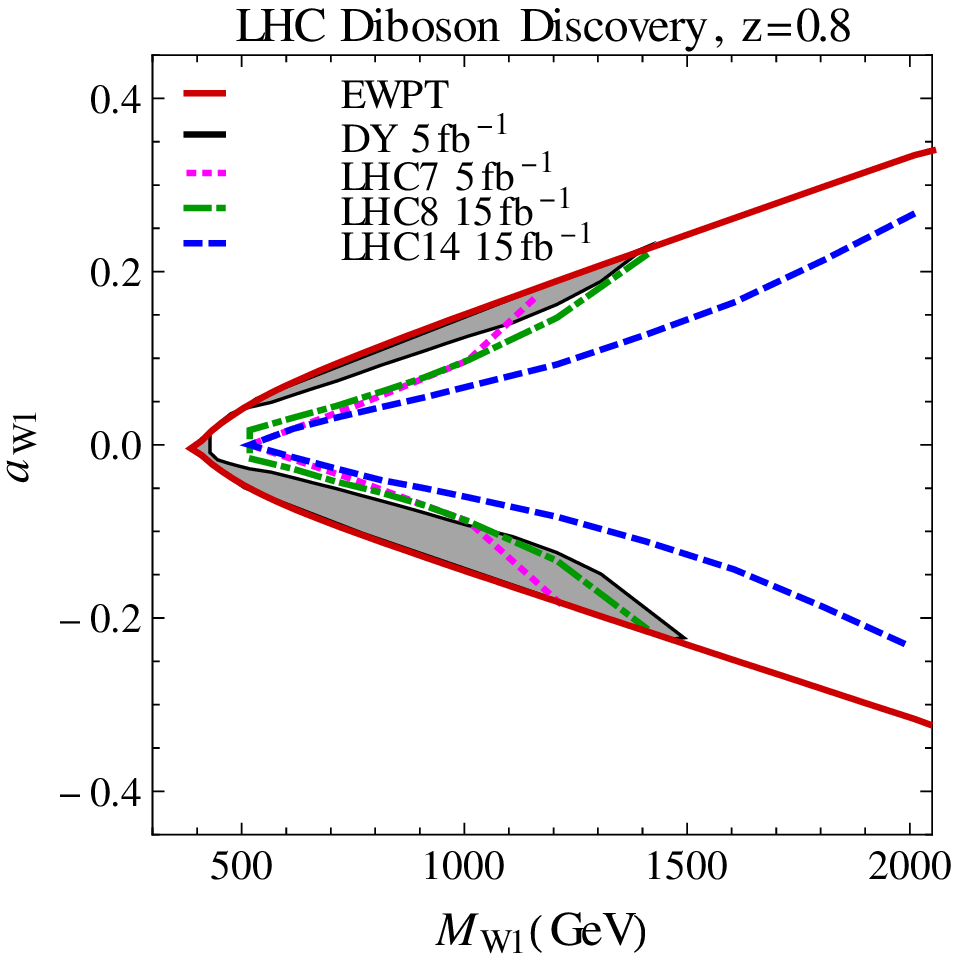,width=7.5cm}}
\put(3.5,-5){\epsfig{file=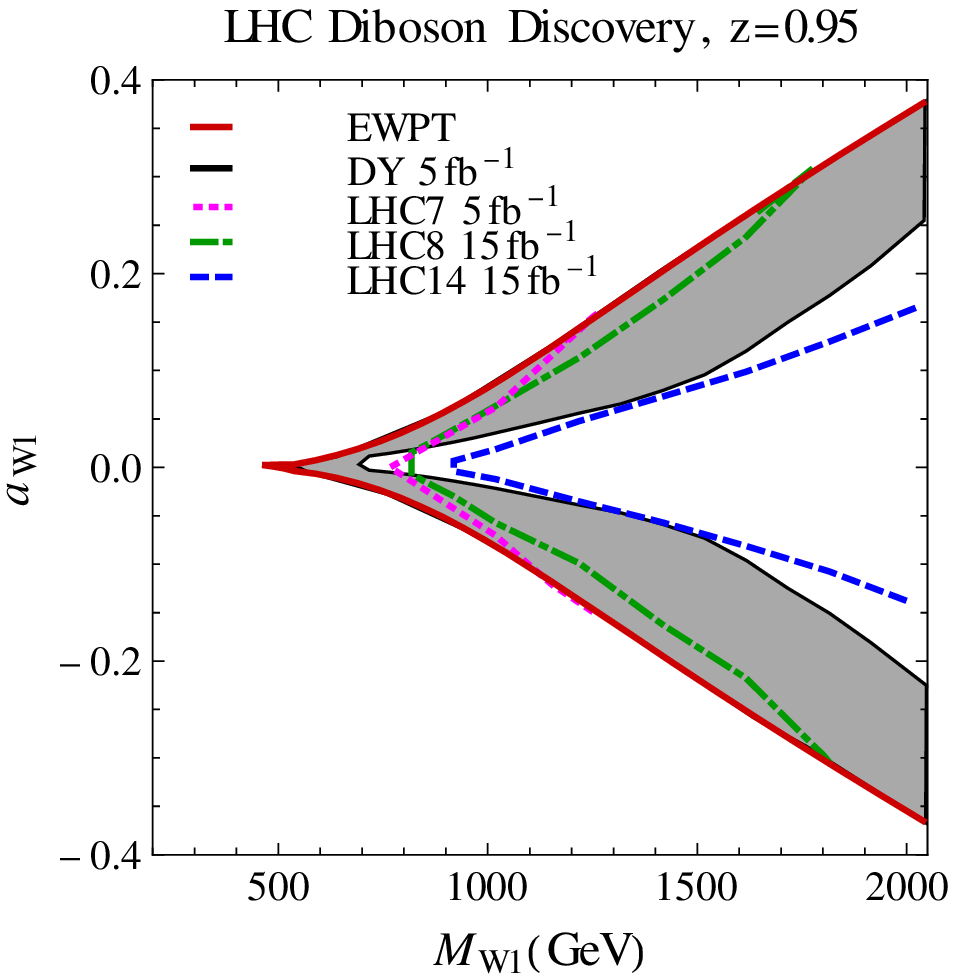,width=7.5cm}}
\end{picture}
\end{center}
\vskip 4.cm \caption{Shown are the $5\sigma$ discovery limits from the LHC on the 4-site model: here, we consider the di-boson channel  for the usual four values of the $z$ parameter. 
We assume all possible LHC setups considered.}
\label{fig:D_dib}
\end{figure}

In this section we compute the actual bounds from the LHC on the 4-site model in considering
 di-boson production, 
and we contrast them to the corresponding figures obtained via (both CC and NC) DY processes.
As explained before, we apply the efficiency on reconstructing the two charged leptons as extracted from the DY channels, supplemented 
by an additional efficiency on the missing energy, and we remind the reader that we made this choice because 
at present there are no published di-boson analyses on possible extra gauge boson
pairs. 
For reference, first we take the current experimental limits as obtained from DY processes (as per previous figures,
hereafter labelled as `DY 5 fb$^{-1}$ in the plots). Then, we consider the 
limits from the following LHC setups assuming di-boson production and decay:
7~TeV with 5~fb$^{-1}$, 
8~TeV with 15~fb$^{-1}$ and
14~TeV with 15~fb$^{-1}$. As before, we consider four representative values of the $z$ parameter.
The results for the exclusion areas are showed in Fig.~\ref{fig:E_dib} whilst those for the
discovery regions are found in Fig.~\ref{fig:D_dib}.  
As we can see from these figures, a large part of the parameter space 
will be explored from the LHC in the next few years, 
excluding or discovering the 4-site model, using the di-boson channel alone.

\begin{figure}[!t]
\begin{center}
\unitlength1.0cm
\begin{picture}(7,10)
\put(-4.3,2.7){\epsfig{file=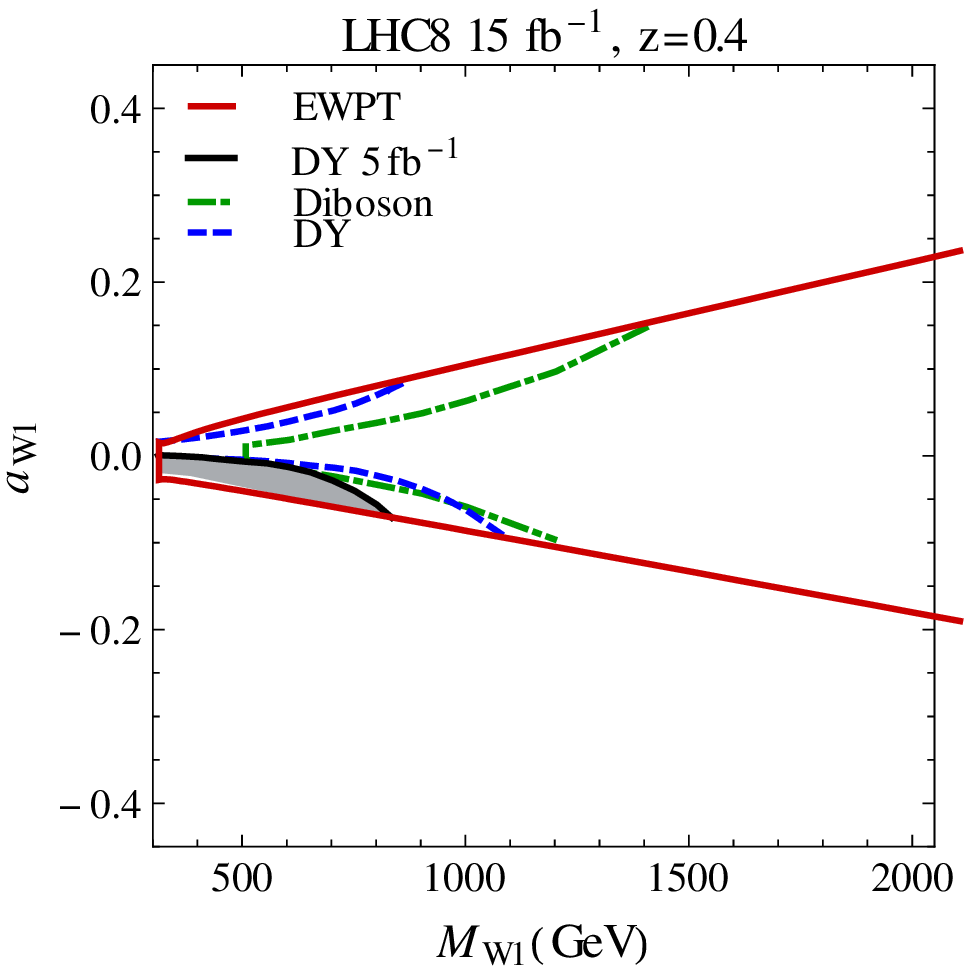,width=7.5cm}}
\put(3.5,2.7){\epsfig{file=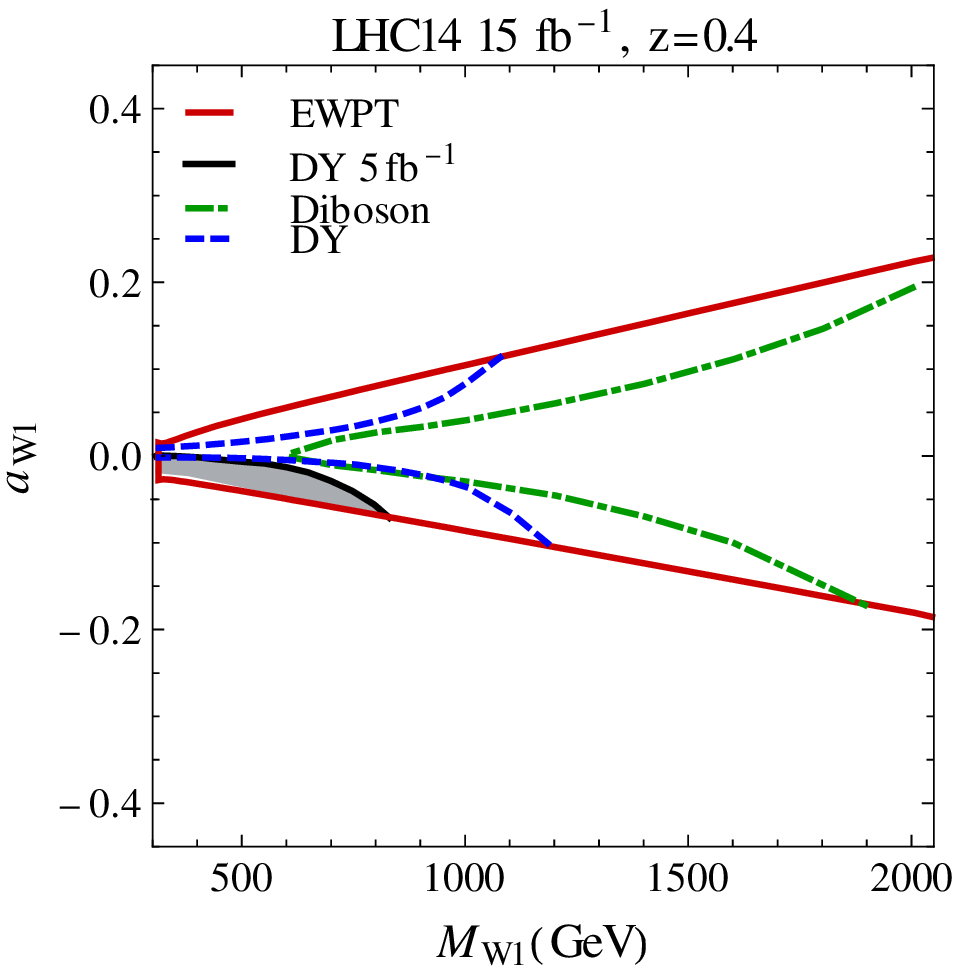,width=7.5cm}}
\put(-4.3,-5){\epsfig{file=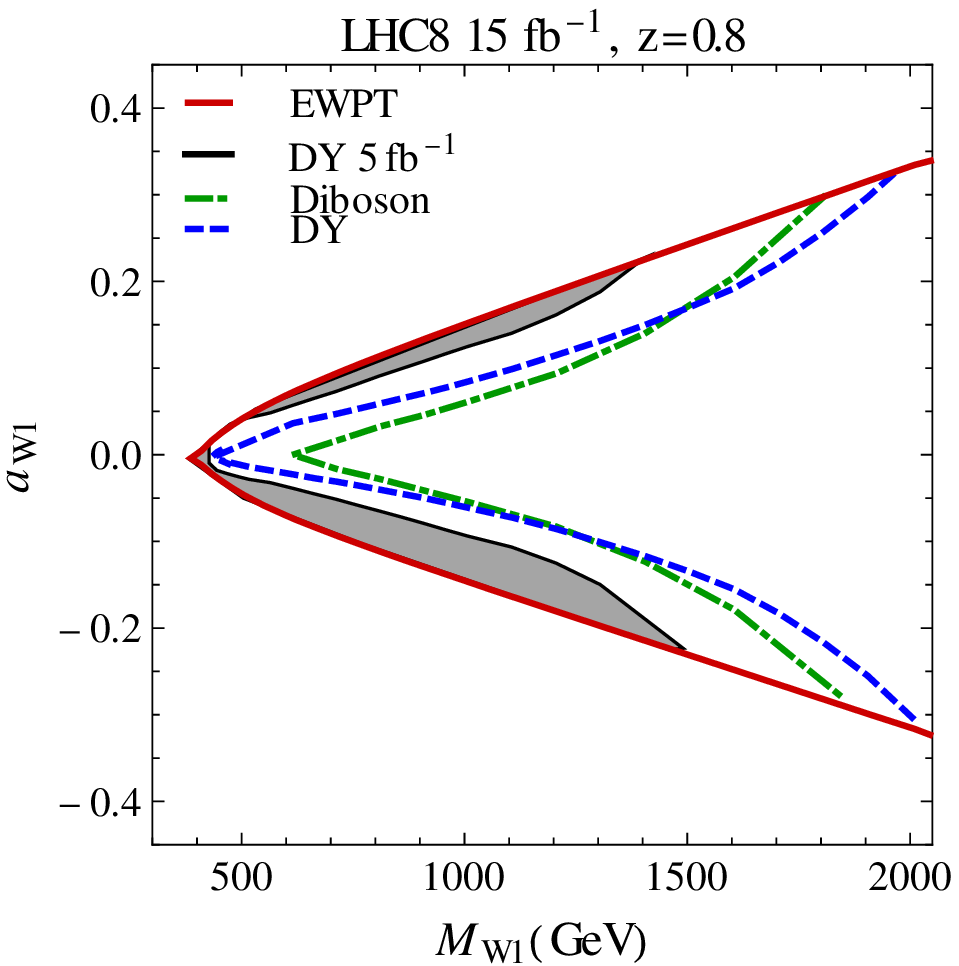,width=7.5cm}}
\put(3.5,-5){\epsfig{file=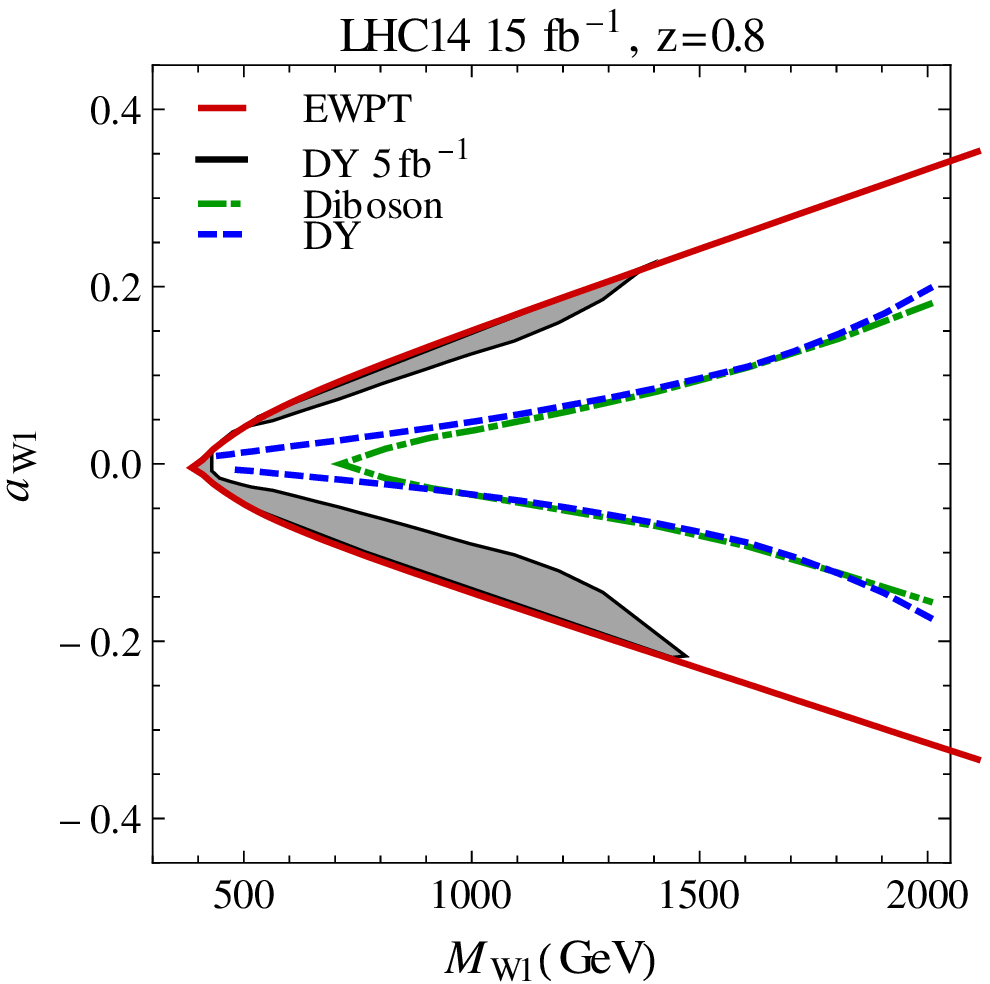,width=7.5cm}}
\end{picture}
\end{center}
\vskip 4.cm \caption{Shown are the 95\% CL exclusion limits from the LHC on the 4-site model: here, we consider both the DY (blue-dashed line) and di-boson (green-dot-dashed line) channels for two values of the $z$ parameter.
We assume only the LHC setups at 8 and 14 TeV with 15 fb$^{-1}$.} 
\label{fig:E+DY}
\end{figure}

Finally, in Fig.~\ref{fig:E+DY} we perform a comparison (using the aforementioned efficiencies) between the di-boson and the DY channels, in 
exclusion limits only, for the LHC at 8 and 14 TeV, both with 15~fb$^{-1}$, for two significant values of $z$. The result is that the di-boson mode is more efficient, both 
at low and high values of the $W_1$ mass, with respect to the DY channels, and this is due to the fact that the trilinear vertex $Z_1WW$ is of the same magnitude as the SM coupling $ZWW$ 
and, moreover,  upon the couplings to the fermions, but only on $z$ and $M_1$, so that the di-boson mode can help exploring also the low coupling 
region.  As we can see from these figures, except for the region of very small $a_{W_1}$ couplings and masses above
1 TeV, the rest of the parameter space which has survived experimental constraints will be explored  from
the LHC in the next few years, 
excluding or discovering the 4-site model, by {synergistically} exploiting both the DY and di-boson channels.

In closing, we should also emphasise that, for reason of space, we have illustrated the scope of DY and di-boson production and decay in setting bounds on our model only limitedly to the case of the charged sector, i.e.,
over the ($M_{W_1}$, $\alpha_{W_1}$) plane.
We can however confirm that a similar pattern can be established in the case of the neutral one as well. i.e., over the  ($M_{Z_1}$, $\alpha_{Z_1}^L(l)$) plane.

\section{Summary and conclusions}

In summary, we have studied the scope of the various LHC stages in testing the parameter space of a 4-site model of strong EWSB 
supplemented with the presence of one composite Higgs boson, compatible with the most recent experimental limits from both EWPTs and direct searches for new Higgs and gauge boson resonances
as well as compliant with theoretical requirements of unitarity.
We have done so by exploiting a process so far largely neglected in experimental analyses, i.e., charged di-boson production into two opposite-charge different-flavour leptons, namely, $e^\pm \mu^\mp E_T^{\rm miss}$ final 
states, where the keyword `charged' refers to the intermediate stage of charged $W$-boson pairs being produced in all combinations possible in our scenario. We then contrasted the yield of this mode with results obtained from
both CC and NC DY processes. In both cases we exploited dedicated parton-level analyses based on acceptance and selection cuts specifically designed  to exalt the complementary role that these two channels
can have at the CERN machine in constraining or revealing our EWSB scenario. Specifically, we have come to the following key conclusions. 
\begin{itemize}
\item DY channels are mostly sensitive to the second gauge boson resonance ($W_2$ in CC and $Z_2$ in NC) whilst charged di-boson production is mostly sensitive to the lightest states, i.e., $W_1$ and $Z_1$, all 
produced in resonant topologies occurring in either subprocess. Therefore, the exploitation of this synergy will eventually enable one to elucidate the full gauge boson spectrum and its dynamics in the context
of our scenario.
\item The di-boson channel, which is entirely new to this study, further offers an advantage over the DY modes, in the sense that it enables one to explore small couplings of the new gauge bosons to the SM fermions,
in virtue of the fact that the overall rate of this process is dependent upon trilinear gauge boson self-couplings which can be very large per se and are further onset in resonant topologies.
\end {itemize}

Benchmark points  of the model under consideration amenable to phenomenogical investigation have been defined and their efficacy in probing different regions of parameter space was emphasised
by adopting all past, current and future setups of the CERN machine.

Finally, a set of numerical tools enabling the accurate prediction of the model spectrum as well as the fast event generation (of both signal and background)
in fully differential form have been produced and are available upon request.

\section*{Acknowledgements}
EA and SM are financed in part through the NExT Institute. LF thanks Fondazione Della Riccia for financial support.

\bibliography{bib_last_2}

\end{document}